\documentclass[preprint,sort&compress,12pt]{elsarticle}

\usepackage{amssymb}
  \usepackage[english]{babel}
  \usepackage{graphics} \graphicspath{{./}{figures/}}
  \usepackage{axodraw}
  \usepackage{rotating}
  \usepackage{url}
  \usepackage{xspace}
  \usepackage{color}
  \usepackage{multirow}
  \usepackage{amsmath}
  \usepackage{comment}

\newcounter{bla}

\journal{Computer Physics Communications}

\usepackage{hyperref}

\newcommand{\Recola}{{\Huge \boldmath{\bf \sc R E C O L A}}}
\newcommand{\recola}{{\sc Recola}}

\newcommand{\pole}{{\sc Pole}}
\newcommand{\openloops}{{\sc OpenLoops}}
\newcommand{\collier}{{\sc Collier}}

\def\mathswitchr#1{\relax\ifmmode{\mathrm{#1}}\else$\mathrm{#1}$\fi}

\newcommand{\PW}{\mathswitchr W}

\newcommand{\PZ}{\mathswitchr Z}

\newcommand{\Pg}{\mathswitchr g}
\newcommand{\PH}{\mathswitchr H}
\newcommand{\Pp}{\mathswitchr p}

\newcommand{\Pu}{\mathswitchr u}
\newcommand{\Pubar}{\bar{\mathswitchr u}}

\newcommand{\Pb}{\mathswitchr b}

\newcommand{\Pep}{\mathswitchr {e^+}}
\newcommand{\Pem}{\mathswitchr {e^-}}

\newcommand{\Pj}{\mathswitchr j}
\newcommand{\Pl}{\mathswitchr \ell}

\newcommand{\m}{\texttt{m}}
\newcommand{\g}{\texttt{g}}

\newcommand{\aaa}{\texttt{a}}
\newcommand{\s}{\texttt{s}}
\newcommand{\n}{\texttt{n}}
\newcommand{\gf}{\texttt{gf}}
\newcommand{\alO}{\texttt{al0}}
\newcommand{\alZ}{\texttt{alZ}}
\newcommand{\als}{\texttt{als}}
\newcommand{\Qren}{\texttt{Qren}}

\newcommand{\Nfren}{\texttt{Nfren}}
\newcommand{\Nf}{\texttt{Nf}}

\newcommand{\dA}{\texttt{d}}
\newcommand{\dB}{\texttt{d2}}
\newcommand{\deltauv}{\texttt{DeltaUV}}
\newcommand{\deltair}{\texttt{DeltaIR}}
\newcommand{\deltairr}{\texttt{DeltaIR2}}
\newcommand{\muuv}{\texttt{muUV}}
\newcommand{\muir}{\texttt{muIR}}

\newcommand{\paA}{\texttt{pa1}}
\newcommand{\paB}{\texttt{pa2}}
\newcommand{\paC}{\texttt{pa3}}
\newcommand{\paD}{\texttt{pa4}}
\newcommand{\fac}{\texttt{fac}}

\newcommand{\npr}{\texttt{npr}}
\newcommand{\processIn}{\texttt{processIn}}
\newcommand{\order}{\texttt{order}}

\newcommand{\gspower}{\texttt{gspower}}

\newcommand{\p}{\texttt{p}}

\newcommand{\res}{\texttt{res}}
\newcommand{\pow}{\texttt{pow}}
\newcommand{\hel}{\texttt{hel}}
\newcommand{\ps}{\texttt{ps}}

\newcommand{\hmu}{\hat{\mu}}
\newcommand{\uv}{{\scriptscriptstyle \rm UV}}
\newcommand{\ir}{{\scriptscriptstyle \rm IR}}
\newcommand{\irr}{{\scriptscriptstyle \rm IR2}}

\newenvironment{remark}{
  \bgroup
  \noindent\unskip\textit{Remark:}
  \ignorespaces}
 {\egroup}

\newenvironment{example}{
  \bgroup
  \noindent\unskip\textit{Example:}
  \par\noindent\unskip
  \ignorespaces}
 {\egroup}

\newlength{\parwidth}
\newenvironment{insertion}
{\setlength{\parwidth}{\textwidth}%
\addtolength{\parwidth}{-\leftmargin}%
\addtolength{\parwidth}{-\labelsep}%
\addtolength{\parwidth}{-3em}%
\\[.3ex]\hspace*{3ex}\begin{minipage}[t]{\parwidth}}
{\end{minipage}\\[.2ex]}

\def\bei{\begin{itemize}}
\def\eei{\end{itemize}}
\def\bq{\begin{equation}}
\def\eq{\end{equation}}
\def\bqa{\begin{eqnarray}}
\def\eqa{\end{eqnarray}}
\newcommand{\ba}[1]{\begin{array}{#1}}
\newcommand{\ea}{\end{array}}

\def\refse#1{\mbox{Section~\ref{#1}}}
\def\refses#1{\mbox{Sections~\ref{#1}}}

\def\citere#1{\mbox{Ref.~\cite{#1}}}
\def\citeres#1{\mbox{Refs.~\cite{#1}}}

\begin{document}

\begin{frontmatter}

\title{\Recola\\
REcursive Computation of One-Loop Amplitudes
\tnoteref{mytitlenote}
}
\tnotetext[mytitlenote]{The program is available from  
  \href{http://recola.hepforge.org/}{\mbox{http://recola.hepforge.org}}.}

\author[a]{Stefano~Actis}
\ead{stefano.actis@gmail.com}
\author[b]{Ansgar~Denner}
\ead{ansgar.denner@physik.uni-wuerzburg.de}
\author[c]{Lars~Hofer}
\ead{hofer@ecm.ub.edu}
\author[b]{Jean-Nicolas~Lang}
\ead{jlang@physik.uni-wuerzburg.de}
\author[b]{Andreas~Scharf}
\ead{ascharf@physik.uni-wuerzburg.de}
\author[d]{Sandro~Uccirati}
\ead{uccirati@to.infn.it}

\address[a]{Lilienstrasse 7, 5200 Brugg AG, Switzerland}
\address[b]{Universit\"at W\"urzburg, 
Institut f\"ur Theoretische Physik und Astrophysik, \\
D-97074 W\"urzburg, Germany}
\address[c]{Department de F\'isica Qu\`antica i Astrof\'isica (FQA), \\
Institut de Ci\`encies del Cosmos (ICCUB),
Universitat de Barcelona (UB), \\
Mart\'i Franqu\`es 1,
E-08028 Barcelona, Spain}
\address[d]{Universit\`a di Torino e INFN, 10125 Torino, Italy}

\begin{abstract}
  We present the {\sc Fortran95} program \recola\ for the perturbative
  computation of next-to-leading-order transition amplitudes in the
  Standard Model of particle physics.  The code provides numerical
  results in the 't~Hooft--Feynman gauge. It uses the complex-mass
  scheme and allows for a consistent isolation of resonant
  contributions.  Dimensional regularization is employed for
  ultraviolet and infrared singularities, with the alternative
  possibility of treating collinear and soft singularities in mass
  regularization.  \recola\ supports various renormalization schemes
  for the electromagnetic and a dynamical $N_{\mathrm{f}}$-flavour
  scheme for the strong coupling constant.  The calculation of
  next-to-leading-order squared amplitudes, summed over spin and
  colour, is supported as well as the computation of colour- and
  spin-correlated leading-order squared amplitudes needed in the
  dipole subtraction formalism.%
\end{abstract}


\end{frontmatter}

\clearpage

\def\thefootnote{\arabic{footnote}}
\setcounter{footnote}{0}




\noindent
{\bf PROGRAM SUMMARY}

\begin{small}
\noindent
{\em Manuscript Title:}  \recola: {REcursive Computation of One-Loop Amplitudes} \\
{\em Authors:} Stefano Actis, Ansgar Denner, Lars Hofer, Jean-Nicolas
Lang, Andreas Scharf, Sandro Uccirati    \\
{\em Program Title: }\recola                                          \\
{\em Journal Reference:}                                      \\
{\em Catalogue identifier:}                                   \\
{\em Licensing provisions:}         GNU GPL version 3
\\
{\em Programming language:} {\sc Fortran95}
\\
{\em Computer:} any with a {\sc Fortran95} compiler                             \\
{\em Operating system:}      Linux, Mac~OS~X  
                             \\
{\em RAM:} Depends on the nature of the problem, typically 1$\,$GB for
a $2\to4$ process.
\\
{\em Number of processors used:}     one                         \\
{\em Supplementary material:}   none 
                              \\
\\
{\em Classification:}\\                 
 4.4 Feynman diagrams,  11.1 General, High Energy Physics and Computing\\
{\em External routines/libraries:}        
\collier\ library
       \\
{\em Subprograms used:}                 none
                       \\
{\em Nature of problem:} \\
Evaluation of general tree-level and one-loop scattering amplitudes
occurring in the calculation of observables in relativistic quantum
field theories
\\
{\em Solution method:}\\
Tree-level and one-loop amplitudes are numerically calculated using a
recursive algorithm. For one-loop amplitudes numerical results for
tensor integrals are needed as input. These are provided by the
\collier\ library. In addition, contributions of
counterterms and rational terms are determined via dedicated Feynman
rules.
   \\
{\em Restrictions:}\\
The code has been used for processes with up to 7 external particles
at one-loop level and up to 9 external particles at tree level. For 
large multiplicities available internal storage may cause limitations.
  \\
{\em Additional comments:}       none
 \\
{\em Running time:}
Depends on the nature of the problem, typically 100$\,$ms for a
$2\to4$ process.

\end{small}

\clearpage

\section{Introduction}
\label{introduction}

The experimental studies at present and future high-energy colliders
are focused on the precise determination of the free parameters of the
Standard Model (SM) and on the search for new physics.  The
interpretation of the data often relies on accurate theoretical
predictions based on perturbation theory, requiring detailed
calculations beyond the leading-order (LO) approximation.  In the past
years, many groups have concentrated their efforts on
next-to-leading-order (NLO) calculations (see e.g.\ 
\citeres{Campbell:2013qaa,Andersen:2014efa,Bern:2008ef,
  Binoth:2010nha,AlcarazMaestre:2012vp}), and alternative strategies
to the traditional Feynman-diagrammatic approach have been developed,
which helped to automatize and speed up the calculation of NLO
amplitudes.  One class of methods makes use of generalized unitarity
relations or of amplitude reduction at the integrand level in order to
directly express one-loop amplitudes in terms of scalar
integrals~\cite{Bern:1994zx,Bern:1994cg,Britto:2004nc,Ossola:2006us,
  Ellis:2007br,Giele:2008ve,Ellis:2008ir,vanHameren:2009dr}.  Other
methods instead rely on higher-rank tensor integrals, either via an
improved diagrammatic approach~\cite{Cascioli:2011va} or employing
one-loop recursion relations~\cite{vanHameren:2009vq,Actis:2012qn}.
Finally, yet another strategy consists in performing a simultaneous 
numerical integration over the phase space and the loop momentum
of NLO amplitudes \cite{Nagy:2003qn,Becker:2010ng,Duplancic:2016lzh}.

The traditional as well as the new techniques for the calculation of
one-loop amplitudes have been implemented in many codes such as {\sc
  FeynArts/ FormCalc}~\cite{Hahn:2000kx,Agrawal:2012cv,Nejad:2013ina},
{\sc CutTools}~\cite{Ossola:2007ax}, {\sc
  Blackhat}~\cite{Berger:2008sj}, {\sc
  Helac-1loop}~\cite{vanHameren:2009dr}, {\sc
  NGluon}~\cite{Badger:2010nx}, {\sc Samurai}~\cite{Mastrolia:2010nb},
{\sc Madloop}~\cite{Hirschi:2011pa}, {\sc GoSam}~\cite{Cullen:2011ac},
and {\sc OpenLoops}~\cite{Cascioli:2011va}.  In this article we
present the {\sc Fortran95} library \recola\ for the generation of
tree-level and one-loop amplitudes in the SM. While almost all of the
above-listed programs were developed with a focus on QCD
corrections\footnote{{\sc FeynArts/FormCalc} allows to perform
  calculations in more general scenarios in and beyond the SM, though
  with less emphasis on high multiplicities and CPU performance.},
\recola\ has been designed from the beginning with the main objective
of facilitating an automated calculation of electroweak (EW)
corrections. Recently, EW corrections have been included also in {\sc
  OpenLoops}~\cite{Kallweit:2014xda,Kallweit:2015dum} and {\sc
  MadGraph5\_aMC@NLO}~\cite{Frixione:2014qaa,Frixione:2015zaa}.
\recola\ further differs from other public codes in the implemented
method as it makes consequent use of a recursive construction of
one-loop off-shell currents following the technique described
in~\citere{Actis:2012qn}.  It has successfully applied for the
calculation of EW corrections to the processes
$\Pp\Pp\to2\Pl{+}{\le}2\Pj$ \cite{Denner:2014ina} and
$\Pp\Pp\to\mu^+\mu^-\Pep\Pem$~\cite{Biedermann:2016yvs}, and for the
calculation of QCD corrections to the process
$\Pp\Pp\to\PW\PW\Pb\bar\Pb\PH$~\cite{Denner:2015yca}.

For the evaluation of the one-loop integrals, a task that demands high
standards with respect to numerical stability and CPU performance,
amplitude generators are either equipped with own internal
implementations, or they rely on external libraries like {\sc FF}
\cite{vanOldenborgh:1990yc}, {\sc LoopTools}~\cite{Hahn:1998yk}, {\sc
  QCDLoop}~\cite{Ellis:2007qk}, {\sc
  OneLOop}~\cite{vanHameren:2010cp}, {\sc
  Golem95C}~\cite{Cullen:2011kv}, {\sc PJFry}~\cite{Fleischer:2010sq},
{\sc Package-X}~\cite{Patel:2015tea}, and
\collier~\cite{Denner:2014gla,Denner:2016kdg}.
In the case of \recola, the public {\sc Fortran95} library
\collier\ is used which achieves a fast
and stable calculation of tensor integrals via the strategies
developed in \citeres{Denner:1999gp,Denner:2005fg,Denner:2010tr}.

The available one-loop generators do not only differ in the class of
computations they can perform, but also in the level of automation and
in the cost of performance (speed and memory). The latter is an
essential aspect because typical Monte-Carlo simulations require a
huge number of evaluations of the matrix element for each partonic
process in order to obtain a sufficient statistical accuracy. To this
end, in the development of \recola\ a big effort has been invested in
the optimization of the performance in order to permit the fast
``on-the-fly'' generation and evaluation of NLO matrix elements. This
strategy is complementary to the one used by other groups, as for
example the {\sc Blackhat} collaboration, who have developed a
flexible storage format of pre-calculated matrix elements for partonic
events in large {\sc Root} $N$-tuple files~\cite{Bern:2013zja}, which
are then read by the Monte Carlo generator.

This article is organized as follows: In \refse{what-is-recola} we
describe the basic features of \recola; \refse{installation} gives the
user the necessary information on how to download and install the
\recola\ library.  \refse{calling recola} explains the usage of
\recola\ and gives a detailed description of its input parameters and
of all subroutines that can be called by the user.  Finally, in
\refse{checks} we list the processes that have been checked against
other programs and we conclude in \refse{conclusions}.

\section{Basic features of \recola}
\label{what-is-recola}

\recola\ is a {\sc Fortran95} code for the computation of
tree-level and one-loop scattering amplitudes in the SM, based 
on recursion relations \cite{Actis:2012qn}. 

The algorithm to compute the tree-level amplitude ${\cal A}_0$ is
inspired by the Dyson--Schwinger
equations~{\cite{Dyson:1949ha,Schwinger:1951ex,Schwinger:1951hq}}.
The recursion relations for the one-loop amplitudes are more involved
and rely on the decomposition of the one-loop amplitude ${\cal A}_1$
in terms of tensor integrals (TIs) $T_{(t)}^{\mu_1\cdots\mu_{r_t}}$
and tensor coefficients (TCs) $c_{\mu_1\cdots\mu_{r_t}}^{(t)}$:
\bq
{\cal A}_1 = 
\sum_{t}\,c_{\mu_1\cdots\mu_{r_t}}^{(t)}\,T_{(t)}^{\mu_1\cdots\mu_{r_t}}
+ {\cal A}_{\rm CT}\,.
\eq
Here, ${\cal A}_{\rm CT}$ is the contribution from the counterterms.
In order to regularize ultraviolet (UV) singularities the TIs are
treated in dimensional regularization by introducing the variable
space-time dimension ${\rm D} = 4 - 2\epsilon$ together with the mass
scale $\mu$:
\bq
T_{(t)}^{\mu_1\cdots\mu_{r_t}} =
\frac{(2\pi\mu)^{4-{\rm D}}}{{\rm i}\pi^2}
\int
\mathrm{d}^{\rm D}\!q\,\frac{q^{\mu_1}\cdots q^{\mu_{r_t}}}{D_{0}^{(t)}\cdots D_{k_t}^{(t)}}.
\eq
Here, $k_t$ is the number of propagators in the loop, $r_t$ the rank
of $T_{(t)}$ and \bq D_i^{(t)} = (q+p_i^{(t)})^2 -( m_i^{(t)})^2,
\qquad i=0,\ldots,k_t, \qquad p^{(t)}_0=0.  \eq UV singularities of
the TIs manifest themselves as poles in $\epsilon$, and they are
cancelled by analogous singularities present in the counterterm
amplitude ${\cal A}_{\rm CT}$, which can be built from tree-level
topologies involving counterterm vertices~\cite{Denner:1991kt}.

Based on an idea by van~Hameren~\cite{vanHameren:2009vq}, a recursive
procedure to compute the TCs numerically has been
developed~\cite{Actis:2012qn} and implemented in \recola .  In this
framework the indices $\mu_1,\dots,\mu_{r_t}$ are taken strictly
4-dimensional (with values $0,1,2,3$), and the $(D-4)$-dimensional
part of the contraction between the TCs and TIs in ${\cal A}_1$ is
taken into account in the form of an additional rational part ${\cal
  A}_{\rm R2}$ (of type $R_2$ \cite{Ossola:2008xq}):
\bq
{\cal A}_1 = 
{\cal A}_{\rm D4} + {\cal A}_{\rm R2} + {\cal A}_{\rm CT}\,,
\qquad
{\cal A}_{\rm D4} = 
\sum_{t}\,c_{\hmu_1\cdots\hmu_{r_t}}^{(t)}\,T_{(t)}^{\hmu_1\cdots\hmu_{r_t}}\,.
\eq
The hat on the indices $\mu_1,\dots,\mu_{r_t}$ indicates that they run
over the 4 dimensions $\hat{\mu}_i=0,1,2,3$.  The tensor integrals
$T_{(t)}^{\hmu_1\cdots\hmu_{r_t}}$ are computed in \recola\ by means
of an interface with the \collier\ library~\cite{Denner:2014gla}.  The
contribution ${\cal A}_{\rm R2}$ is determined evaluating
tree-level-like topologies with special Feynman
rules~\cite{Ossola:2008xq,Draggiotis:2009yb,Garzelli:2009is,Shao:2011tg}
similar to those for the counterterms.

\subsection{Collinear and soft singularities}
\label{collinear and soft}

Collinear singularities originating from light fermions can be treated
in \recola\ either in dimensional or in mass regularization.  For each
fermion an individual choice can be made: If a fermion is defined as
massless, collinear divergences stemming from this fermion are
regularized dimensionally. If on the other hand it has been assigned a
(regulator) mass, its collinear singularities are regularized by the
corresponding mass parameter.

Soft singularities are either regularized dimensionally or by
assigning a mass regulator $\lambda$ to photons and gluons.  The
second case is allowed in \recola\ only if collinear singularities are
treated in mass regularization for all fermions.

\subsection{Dimensional regularization}
\label{dimensional regularization}

If dimensional regularization is used for collinear or soft
singularities, poles in $\epsilon$ of infrared (IR) origin are
generated in ${\cal A}_1$ together with a dependence on the scale
$\mu$.  In order to distinguish this $\epsilon$ and $\mu$~dependence
of IR origin from the one of UV origin, \recola\ introduces the
separate parameters $\mu_\uv$, $\epsilon_\uv$ and $\mu_\ir$,
$\epsilon_\ir$ in all TIs and counterterms, together with
\bqa
\Delta_\uv &=& 
\frac{(4\pi)^{\epsilon_\uv}\,\Gamma(1+\epsilon_\uv)}{\epsilon_\uv},
\nonumber\\
\Delta_\ir &=& 
\frac{(4\pi)^{\epsilon_\ir}\,\Gamma(1+\epsilon_\ir)}{\epsilon_\ir},
\qquad
\Delta_\irr \;=\; 
\frac{(4\pi)^{\epsilon_\ir}\,\Gamma(1+\epsilon_\ir)}{\epsilon_\ir^2}.
\eqa
Following the conventions of
\collier~\cite{Denner:2014gla,Denner:2016kdg}
and \citere{Denner:2010tr}, the parameters $\Delta_\uv$, $\Delta_\ir$
and $\Delta_\irr$ that contain the poles in $\epsilon$ absorb a
normalization factor of the form $1+{\cal O}(\epsilon)$.  In terms of
these parameters, the one-loop amplitude takes the general form
\bq
{\cal A}_1 = 
  \Delta_\uv\,{\cal A}_1^\uv
+ \Delta_\irr\,{\cal A}_1^\irr
+ \Delta_\ir\,{\cal A}_1^\ir(\mu_\ir)
+ {\cal A}_1^{\rm fin}(\mu_\uv,\mu_\ir).
\eq
The term ${\cal A}_1^\uv$ vanishes after renormalization.  The a
priori unphysical scale $\mu_\uv$ should either cancel between the TIs
and the counterterms, or it should receive a physical interpretation
as for example when it is identified with the renormalization scale
$Q$ in the $\overline{\rm MS}$ scheme for the strong coupling constant
$g_s$.  In \recola, however, the $\overline{\rm MS}$ renormalization
scale $Q$ is kept independent from $\mu_\uv$.  In this way, also
$\overline{\rm MS}$-renormalized amplitudes are independent of
$\mu_\uv$ but depend on $Q$ instead.  

The counterterm $\delta Z_{g_s}$, relating the bare and the
renormalized strong coupling constants $g_s^0$ and $g_s$ according to
$g_s^0 = g_s \, ( 1 + \delta Z_{g_s} )$ is defined as
\bq
\delta Z_{g_s} = 
- \,\frac{\alpha_s(Q^2)}{4\pi}\,\Biggl[\,
    \left( \frac{11}{2} - \frac{N_{\mathrm{f}}}{3} \right)
    \left( \Delta_\uv + \ln\frac{\mu_\uv^2}{Q^2} \right)
  - \frac{1}{3}\sum_{F}
    \left( \Delta_\uv + \ln\frac{\mu_\uv^2}{m_{F}^2} \right)
  \,\Biggr],
  \label{eq:Zgs}
\eq
where $N_{\mathrm{f}}$ is the number of active (light) flavours and
$F$ runs over the inactive (heavy) flavours.  
According to eq.~(\ref{eq:Zgs}),
the contribution from active flavours is renormalized within the
$\overline{\rm MS}$ scheme, while the one from inactive flavours is
subtracted at zero momentum transfer.
The classification into
\textit{active} and \textit{inactive} flavours defines the flavour
scheme. In \recola\ the user can choose between: 
\bei
\item The \textit{variable-flavour scheme}: All quark flavours lighter
  than $Q$ are considered as active, the remaining ones are treated as
  inactive.
\item The \textit{$N_{\mathrm{f}}$-flavour scheme}: The
  $N_{\mathrm{f}}$ lightest quarks are considered active, the
  remaining ones are treated as inactive. In this case,
  $N_{\mathrm{f}}$ cannot be chosen lower than the number of massless
  quarks.  
\eei 
After renormalization, the one-loop amplitude becomes
\bq
{\cal A}_1 = 
  \Delta_\irr\,{\cal A}_1^\irr
+ \Delta_\ir\,{\cal A}_1^\ir(\mu_\ir)
+ {\cal A}_1^{\rm fin}(Q,\mu_\ir).
\eq
\recola\ performs a numerical computation of the complete amplitude
${\cal A}_1$, with the values for $\Delta_\ir$, $\Delta_\irr$,
$\mu_\ir$ and $Q$ supplied by the user.  The contribution ${\cal
  A}_1^{\rm fin}$ can be obtained by setting $\Delta_\ir = \Delta_\irr
= 0$ (default). Since the computation of ${\cal A}_1$ involves objects
depending on $\Delta_\uv$ and $\mu_\uv$ at intermediate steps,
numerical values for these variables must be given as well.  The
independence of ${\cal A}_1$ on $\Delta_\uv$ and $\mu_\uv$ can be
verified numerically by varying these parameters.  \recola\ provides
default values for all the above-mentioned parameters that can be
changed by the user.

\subsection{The strong coupling constant $\alpha_s$}
\label{alphas}

The renormalized strong coupling constant $\alpha_s$ depends on the
renormalization scale $Q$.  It is appropriate to choose a value for
$Q$ of the order of the energy scale characteristic for the process in
question and to take as coupling constant the corresponding value for
$\alpha_s(Q^2)$.

Often, in the computation of physical processes the scale $Q$ is
defined from the momenta of the external particles and is thus
assigned a different value for each phase-space point. The possibility
to use such a dynamical scale $Q$ is supported by \recola, and the
respective values for the running $\alpha_s(Q^2)$ can either be
supplied by the user or computed by \recola.  In the latter case, the
value for $\alpha_s(Q^2)$ at the scale $Q$ is determined from its
value $\alpha_s(Q_0^2)$ at the initialization scale $Q_0$, by means of
the following one- and two-loop formulas in the
$N_{\mathrm{f}}$-flavour scheme~\cite{Ellis:1991qj}:
\bqa
\mbox{1-loop running:}
&&
a = \frac{a_0}{1 + a_0\,\beta_0\,L},
\\
\mbox{2-loop running:}
&&
a = 
a_0\,\bigg[ 
  1 
+ a_0\,\beta_0\,L 
+ a_0\,b_1\ln\left( 1 + \frac{a_0\,\beta_0\,L}{1+a_0\,b_1} \right) 
\bigg]^{-1}\!.\;\;\;\;
\eqa
Here, we have introduced
\bq
a = \frac{\alpha_s(Q^2)}{4\pi},
\qquad
a_0 = \frac{\alpha_s(Q_0^2)}{4\pi},
\qquad
L = \ln\frac{Q^2}{Q_0^2},
\eq
\bq
b_1 = \frac{\beta_1}{\beta_0},
\qquad
\beta_0 = 11 - \frac{2}{3}N_{\mathrm{f}},
\qquad
\beta_1 =  102 - \frac{38}{3}N_{\mathrm{f}}.
\eq

\subsection{Electroweak renormalization}
\label{ew renormalization}

Due to the presence of massive and unstable gauge bosons in the
EW sector of the SM, complex masses have to be
introduced without spoiling gauge invariance.  This is achieved in
\recola\ by using the complex-mass scheme
of~\citeres{Denner:1999gp,Denner:2005fg,Denner:2006ic}.  The user can,
however, also choose to proceed in the on-shell scheme, where only the
real part of the self-energies is taken for the computation of the
counterterms and the imaginary part of the masses is kept only in the
denominator of propagators.

For the renormalization of the EW coupling constant $\alpha$, 
the user can choose between three different schemes: 
\bei
\item $G_{\rm F}$ scheme:\\
  In this scheme the renormalized electromagnetic coupling $\alpha$ is
  derived from the Fermi constant $G_{\rm F}$, measured in muon decay, 
  and the masses of the $\PW$ and $\PZ$~bosons
  via the tree-level relation:
\bq
\alpha = 
\frac{\sqrt{2}\,G_{\rm F}}{\pi}\,{\rm Re}(M_{\rm W}^2)\,
\left( 1 - \frac{{\rm Re}(M_{\rm W}^2)}{{\rm Re}(M_{\rm Z}^2)} \right).
\eq
\item $\alpha(0)$ scheme:\\
In this scheme $\alpha$ is fixed from the value measured in
Thomson scattering at $p^2 = 0$.
\item $\alpha(M_{\PZ})$ scheme:\\
  In this scheme $\alpha$ is renormalized at the $\PZ$ pole, thus
  implicitly taking into account its running from $p^2 = 0$ to $p^2 =
  M_{\rm Z}^2$.
\eei

\subsection{Amplitude structure}
\label{amplitude structure}

In general, 
the amplitude of a process depends on helicities and 
colours of external particles. 
For a process with $l$ external legs
we introduce the notation
\bq
{\cal A} \to {\cal A}_{a_1 \cdots a_l}[h_1,\dots,h_l].
\eq
The variable $h_n$ ($n=1,\dots,l$) defines the helicity of the 
$n^{\rm th}$ external particle and can take the values $+$ or $-$ for 
fermions and massless vector bosons, and the values $+$, $-$ or $0$ for 
massive vector bosons
(see \refse{conventions} for the conventions used in \recola). 
For scalar particles it necessarily assumes the value $0$.
The variable $a_n$ ($n=1,\dots,l$) defines the colour 
of the $n^{\rm th}$ external particle and can take the values $1,2,3$ 
for quarks and anti-quarks, and the values $1,\dots,8$ for gluons,
while it is absent 
for colourless particles. 

According to the $\rm SU(3)$ decomposition $3\otimes\bar{3}=8\oplus
1$, the colour-octet representation of the gluon can be related to the
product of the fundamental and anti-fundamental representation of
quarks and anti-quarks.  The colour content of the amplitude can thus
alternatively be expressed in the so-called colour-flow
representation, where the colour of the gluons is represented by means
of a pair of indices taking the values $1,2,3$ (see
\citeres{Actis:2012qn,Kanaki:2000ms,Maltoni:2002mq} for details).
Following this approach we write the amplitude as
\bq
{\cal A} \to {\cal A}^{i_1 \cdots i_l}_{j_1 \cdots j_l}[h_1,\dots,h_l].
\eq
The colour index $i_n$ ($n=1,\dots,l$) is absent for colourless
particles, incoming quarks and outgoing anti-quarks, while it takes
the values $1,2,3$ for gluons, incoming anti-quarks and outgoing
quarks.  Similarly, the anti-colour index $j_n$ ($n=1,\dots,l$) is
absent for colourless particles, incoming anti-quarks and outgoing
quarks, while it takes the values $1,2,3$ for gluons, incoming quarks
and outgoing anti-quarks.  The unphysical colour-singlet component,
implicitly contained in the decomposition of the $3\otimes \bar{3}$
representation of the gluons, is eliminated by requiring
\bq
\sum_{i_m,j_m}\delta_{j_m}^{i_m}\,
{\cal A}^{i_1 \cdots i_m \dots i_l}_{j_1 \cdots j_m \dots j_l} = 0,
\eq
if particle $m$ is a gluon.
The two parametrizations 
(colour-octet vs. colour-flow representation of the gluon)
are related through
\bq
{\cal A}^{i_1 \cdots i_l}_{j_1 \cdots j_l}[h_1,\dots,h_l] = 
{\sum_{a_1, \dots, a_l}\!}
{(\Delta_{a_1})^{i_1}}_{j_1}\,\cdots\,{(\Delta_{a_l})^{i_l}}_{j_l}\,
{\cal A}_{a_1 \cdots a_l}[h_1,\dots,h_l],
\eq
\bq
{\cal A}_{a_1 \cdots a_l}[h_1,\dots,h_l] =
{\sum_{\substack{i_1, \dots, i_l\\j_1, \dots, j_l}}\!}
{(\Delta_{a_1})^{j_1}}_{i_1}\,\cdots\,{(\Delta_{a_l})^{j_l}}_{i_l}\,
{\cal A}^{i_1 \cdots i_l}_{j_1 \cdots j_l}[h_1,\dots,h_l], 
\eq
where the sums run over the colour indices present in 
${\cal A}_{a_1 \cdots a_l}$ and ${\cal A}^{i_1 \cdots i_l}_{j_1 \cdots
  j_l}$, respectively.  
The matrices ${(\Delta_a)^{i}}_{j}$ are given by
\bq
{(\Delta_a)^{i}}_{j} =
\left\{
\begin{array}{ll}
\delta_a^{i}
& \quad \mbox{for incoming anti-quarks and} \\
& \quad \mbox{outgoing quarks ($j$ is absent)}
\\[1.2ex]
\delta_{a\,j}
& \quad \mbox{for incoming quarks and} \\
& \quad \mbox{outgoing anti-quarks ($i$ is absent)}
\\[1.2ex]
\frac{1}{\sqrt{2}}{(\lambda_a)^{i}}_{j} & \quad
\mbox{for gluons}
\end{array}
\right.\;,
\eq
where $\lambda_a$ ($a=1,\dots,8$) are the usual Gell-Mann
matrices.\footnote{In our convention, the $\lambda_a$ are normalized
  according to ${\rm Tr}(\lambda_a\lambda_b) = 2\,\delta_{ab}$.}
Note that, while \recola\ exclusively works in the colour-flow
formalism, we give all formulae in both representations to allow
for an easy translation from one to the other.

In the colour-flow representation the colour part of the Feynman rules 
contains products of Kronecker $\delta$s, and the colour structure of the 
amplitude can be obtained as a linear combination of all possible
structures built from
products of Kronecker $\delta$s carrying the colour indices of the 
external particles:
\bq
{\cal A}^{i_1 \cdots i_l}_{j_1 \cdots j_l}[h_1,\dots,h_l] = 
{\sum_{P}}\,
\delta^{i_1}_{j_{\!_{P(1)}}}\,\delta^{i_2}_{j_{\!_{P(2)}}}\,\cdots\,
\delta^{i_l}_{j_{\!_{P(l)}}}\,
{\cal A}_{ {}_{P(1)} {}_{P(2)} \dots {}_{P(l)}}[h_1,\dots,h_l].
\eq
Here, the sum runs over all possible permutations $P$ of the labels of
external gluons, incoming quarks, and outgoing anti-quarks.  The
amplitudes ${\cal A}_{ {}_{P(1)} {}_{P(2)} \dots
  {}_{P(l)}}[h_1,\dots,h_l]$ are called structure-dressed amplitudes.
For example in the process ${\rm u}\,\bar{{\rm u}}\,\to\,\PZ\,{\rm
  g}\,{\rm g}$, this decomposition is given by
\bqa
{\cal A}^{i_2 i_4 i_5}_{j_1 j_4 j_5} 
&=&\phantom{{}+{}}
  \delta^{i_2}_{j_1}\,\delta^{i_4}_{j_4}\,\delta^{i_5}_{j_5}\,
  {\cal A}_{145}\,
+ \delta^{i_2}_{j_1}\,\delta^{i_4}_{j_5}\,\delta^{i_5}_{j_4}\,
  {\cal A}_{154}\,
+ \delta^{i_2}_{j_4}\,\delta^{i_4}_{j_1}\,\delta^{i_5}_{j_5}\,
  {\cal A}_{415}\,
\nonumber\\
&&{}+ \delta^{i_2}_{j_4}\,\delta^{i_4}_{j_5}\,\delta^{i_5}_{j_1}\,
    {\cal A}_{451}\,
+   \delta^{i_2}_{j_5}\,\delta^{i_4}_{j_1}\,\delta^{i_5}_{j_4}\,
    {\cal A}_{514}\,
+   \delta^{i_2}_{j_5}\,\delta^{i_4}_{j_4}\,\delta^{i_5}_{j_1}\,
    {\cal A}_{541},
\eqa
where $j_1$ denotes the colour index of the incoming quark, $i_2$
the one of the incoming anti-quark, and $i_4,j_4$ and $i_5,j_5$
the ones of the outgoing gluons.
In order to render the notation more compact, we introduce the following
$l$-dimensional vectors:
\bei
\item
$\vec{h}\,$ with components $h_1,\dots,h_l$, 
\item
$\vec{c}\,$ with components $c_1,\dots,c_l$, where $c_n=P(n)$ if 
$\delta^{i_n}_{j_{{P(n)}}}$ is present in the colour structure (i.e.\ 
if particle $n$ is a gluon, an incoming anti-quark or an outgoing 
quark) and $c_n = 0$ otherwise
(i.e.\ if particle $n$ is a colourless particle, an incoming quark or an 
outgoing anti-quark).
\eei
We then rewrite the structure-dressed amplitudes in the compact form 
\bq
{\cal A}^{(\vec{c},\vec{h})} = 
{\cal A}_{ {}_{P(1)} {}_{P(2)} \dots {}_{P(l)}}[h_1,\dots,h_l].
\eq
\recola\ computes the Born contribution ${\cal
  A}_0^{(\vec{c},\vec{h})}$ and the one-loop contribution ${\cal
  A}_1^{(\vec{c},\vec{h})}$ to the structure-dressed amplitudes ${\cal
  A}^{(\vec{c},\vec{h})}$ for all values the vectors $\vec{c}$ and
$\vec{h}$ can take according to the external particles of the process.

\subsection{Squared amplitudes}
\label{squared amplitudes}

\recola\ also computes the squared amplitude, summed over helicities
and colours of the outgoing particles and averaged over helicities and
colours of the incoming ones (the average is indicated by a ``bar''):
\bq
\overline{{\cal A}^2} = 
{\overline{\!\!\!\sum_{a_1, \dots, a_l}\!\!\!}\;}\;\;\;
{\overline{\!\!\!\sum_{h_1, \dots, h_l}\!\!\!}\;}\;\;
\Big|{\cal A}_{a_1 \cdots a_l}[h_1,\dots,h_l]\Big|^2 = 
{\overline{\!\!\sum_{\substack{i_1, \dots, i_l\\j_1, \dots, j_l}}\!\!}\;}\;\;\;
{\overline{\!\!\!\sum_{h_1, \dots, h_l}\!\!\!}\;}\;
\Big|{\cal A}^{i_1 \cdots i_l}_{j_1 \cdots j_l}[h_1,\dots,h_l]\Big|^2.
\eq

An order-by-order expansion of the previous formula defines the Born
and one-loop contribution to the squared amplitude.  Omitting for
compactness the colour and helicity indices, we have
\bq
\overline{{\cal A}^2} = 
\overline{\sum}\,\big|{\cal A}_0 + {\cal A}_1 + \dots \big|^2 = 
\overline{\sum}\;\bigg\{ 
  |{\cal A}_0|^2 
+ 2\,{\rm Re}\,({\cal A}_1\,{\cal A}_0^*)
+ |{\cal A}_1|^2
+ \dots
\bigg\}.
\eq
At the Born level 
\recola\ computes the squared amplitude 
$\big(\,\overline{\!{\cal A}^2\!}\,\big)_{\!0}$, given by
\bq
\big(\,\overline{\!{\cal A}^2\!}\,\big)_{\!0} = \overline{\sum}\,|{\cal A}_0|^2,
\eq
at the one-loop level it 
computes the one-loop contribution 
$\big(\,\overline{\!{\cal A}^2\!}\,\big)_{\!1}$ 
to the squared amplitude. 
If the Born amplitude ${\cal A}_0$ 
does not vanish, 
$\big(\,\overline{\!{\cal A}^2\!}\,\big)_{\!1}$ is calculated as
\bq
\big(\,\overline{\!{\cal A}^2\!}\,\big)_{\!1} =
\overline{\sum}\,2\,{\rm Re}\,({\cal A}_1\,{\cal A}_0^*).
\eq
For processes with a vanishing Born amplitude ${\cal A}_0$ (but non-vanishing 
${\cal A}_1$), \recola\ computes the first non-vanishing contribution 
to the squared amplitude, which in this case amounts to
\bq
\big(\,\overline{\!{\cal A}^2\!}\,\big)_{\!1} = 
\overline{\sum}\,|{\cal A}_1|^2 .
\eq
\subsection{Colour- and spin-correlated squared amplitudes}
\label{correlation}

In order to compute the subtraction terms in the Catani--Seymour
dipole
formalism~\cite{Catani:1996vz,Catani:2002hc}, colour- and
spin-correlated squared tree-level
amplitudes are needed.

Being ${\cal A}_{a_1 \cdots a_l}$ the amplitude of a
process, one builds for every coloured particle $n=1,\dots,l$ the
amplitude ${\cal A}_{a_1 \cdots a_l\,a}(n)$, where the original colour
structures have been extended by an additional factor describing the
emission of a gluon of colour $a$ from particle $n$ of the
original process:
\bq
{\cal A}_{a_1 \cdots a_n \cdots a_l\,a}(n) =
\sum_{a_n'}\,(T_a)_{a_n\,a_n'}\,{\cal A}_{a_1 \cdots a_n' \cdots a_l},
\qquad\text{with}\label{eq:ColCorr1}
\eq
\bq
(T_a)_{a_n\,a_n'} =
\left\{
\begin{array}{ll}
+\,\frac{1}{\sqrt{2}}(\lambda_a)_{a_n\,a_n'}
& \quad
\mbox{for incoming anti-quarks}\\
& \quad \mbox{and outgoing quarks}
\\[1.2ex]
-\,\frac{1}{\sqrt{2}}(\lambda_a)_{a_n'\,a_n}
& \quad
\mbox{for incoming quarks}\\
& \quad \mbox{and outgoing anti-quarks}
\\[1.2ex]
{\rm i}\,f_{a_n\,a\,a_n'}
& \quad
\mbox{for gluons}
\end{array}
\right.\;.\label{eq:ColCorr2}
\eq
Here, $f_{a\,b\,c}$ are the structure constants\footnote{The
  normalization for $f_{abc}$ is fixed by $[T_a,T_b] = {\rm
    i}\,f_{abc}\,T_c$.}  of $\mathrm{SU}(N_{\mathrm{c}})$ and
$N_{\mathrm{c}}=3$. In the colour-flow formalism used by \recola,
Eqs.~(\ref{eq:ColCorr1}) and (\ref{eq:ColCorr2}) translate into 
\bq
{\cal A}^{i_1 \cdots i_n \cdots i_l\,i}_{j_1 \cdots j_n \cdots
  j_l\,j}(n) = \sum_{i_n',j_n'}\, K_{j\,j_n i_n'}^{i\,i_n j_n'}\,
{\cal A}^{i_1 \cdots i_n' \cdots i_l}_{j_1 \cdots j_n' \cdots j_l},
\eq
and
\bq
K_{j\,j_n i_n'}^{i\,i_n j_n'} =
\left\{
\begin{array}{ll}
+\,\delta^{i_n}_{j}\delta^{i}_{i_n'} 
- \frac{1}{N_{\mathrm{c}}}\delta^{i_n}_{i_n'}\delta^{i}_{j}
& \quad\mbox{for incoming anti-quarks and out-}\\
& \quad\mbox{going quarks ($j_n$ and $j_n'$ are absent)}
\\[1.2ex]
-\,\delta^{j_n'}_{j}\delta^{i}_{j_n} 
+ \frac{1}{N_{\mathrm{c}}}\delta^{j_n'}_{j_n}\delta^{i}_{j}
& \quad\mbox{for incoming quarks and outgoing}
\\ & \quad\mbox{anti-quarks ($i_n$ and $i_n'$ are absent)}
\\[1.2ex]
\delta^{i_n}_{j}\delta^{i}_{i_n'}\delta^{j_n'}_{j_n}
- \delta^{i_n}_{i_n'}\delta^{i}_{j_n}\delta^{j_n'}_{j}
& \quad\mbox{for gluons}
\end{array}
\right., 
\eq 
where ${\cal A}^{i_1 \cdots i_l}_{j_1\cdots j_l}$ denotes
the original amplitude, ${\cal A}^{i_1 \cdots i_l\,i}_{j_1 \cdots
  j_l\,j}(n)$ the modified one, and the additional gluon has colour
indices ($i,j$).

\recola\ computes, at the Born level, the colour-correlated squared 
amplitude $\big(\,\overline{\!{\cal A}^2\!}\,\big)_{\!\mathrm{c}}(n,m)$ 
between particle $n$ and $m$ for all pairs $(n,m)$, defined as
\bqa
\big(\,\overline{\!{\cal A}^2\!}\,\big)_{\!\mathrm{c}}(n,m) 
\!\!&=&\!\! 
c_n
\overline{\sum}
\big({\cal A}_{a_1 \cdots a_l\,a}(n)[h_1,\dots,h_l]\big)^*\,
{\cal A}_{a_1 \cdots a_l\,a}(m)[h_1,\dots,h_l] 
\nonumber\\
\!\!&=&\!\!
c_n
\overline{\sum}
\big({\cal A}^{i_1 \cdots i_l\,i}_{j_1 \cdots j_l\,j}(n)[h_1,\dots,h_l]\big)^*\,
{\cal A}^{i_1 \cdots i_l\,i}_{j_1 \cdots j_l\,j}(m)[h_1,\dots,h_l],\quad
\eqa
where the sum is over all colour and helicity indices.
The global factor $c_n$ is given by
$c_n=1/(2C_{\rm F})$ if particle $n$ is a quark or an anti-quark
and by $c_n=1/(2C_{\rm A})$ if particle $n$ is a gluon.%
\footnote{The Casimir operators in the fundamental and adjoint
  representation are defined as $\sum_a (T_aT_a)_{bc} = 2C_{\rm
    F}\delta _{bc}$ and $\sum_{c,d} f_{acd}f_{bcd} = 2C_{\rm A}\delta
  _{ab}$, respectively, such that $C_{\rm
    F}=(N_{\mathrm{c}}^2-1)/(2N_{\mathrm{c}})$ and $C_{\rm
    A}=N_{\mathrm{c}}$ as usual.}  
These choices are made such that $\big(\,\overline{\!{\cal
    A}^2\!}\,\big)_{\!\mathrm{c}}(n,m)$ is independent of the
normalization chosen for $(T_a)_{a_n\,a_n'}$.

\begin{sloppypar}
  In the case of a ${q}\bar{q}$ or a ${\rm gg}$ splitting, the
  Catani--Seymour subtraction formalism requires also spin
  correlations to build the subtraction terms (the non-diagonal terms
  of Eqs.\ (5.8), (5.9), (5.40), (5.41), (5.67), (5.68), (5.99),
  (5.100), (5.147), (5.148), (5.167), (5.168), (5.185), (5.186)
  of~\citere{Catani:1996vz}).  These terms can essentially be obtained
  replacing the polarization vector of the splitting gluon by an
  appropriate four-vector.  To this end, \recola\ provides, at the
  Born level, the spin--colour-correlated squared amplitude
  $\big(\,\overline{\!{\cal A}^2\!}\,\big)_{\!\mathrm{sc}}(n,m,v)$ for
  pairs $(n,m)$ of external gluons $n$ and external coloured particles
  $m$.  It is given by the colour-correlated squared amplitude
  $\big(\,\overline{\!{\cal A}^2\!}\,\big)_{\!\mathrm{c}}(n,m)$
  computed with the special vector $v$ (to be provided by the user)
  instead of the usual polarization vector for gluon $n$.
\end{sloppypar}

The QCD subtraction formalism of~\citere{Catani:1996vz} can be adapted
for application to QED by replacing gluons with photons and colour
with electric charge.  Due to the Abelian nature of QED, charge
correlation does not involve special amplitudes  (unlike colour
correlation) and can be trivially built from the squared Born
amplitude.  In the subtraction method for QED, spin correlation is
needed in the case of a photon splitting into a fermion anti-fermion
pair, which, in an analogous manner to the QCD case, can be obtained
by replacing the polarization vector of the corresponding photon with
an appropriate four-vector.  For this purpose, \recola\ provides, at
the Born level, the spin-correlated squared amplitude
$\big(\,\overline{\!{\cal A}^2\!}\,\big)_{\!\mathrm{s}}(n,v)$ for
external photons $n$. It is given by the squared amplitude
$\overline{{\cal A}^2}$ computed with a special vector $v$ (to be
provided by the user) instead of the usual polarization vector for
photon~$n$.

\subsection{Selecting intermediate states and resonances}
\label{resonances}

As a further useful feature, \recola\ offers the possibility to
request specific intermediate states in the amplitude, i.e.\ to select
those contributions where the final state of the process is reached
via one or more definite intermediate particles.  The cross-section
of a process is often dominated by contributions where these
intermediate states become resonant. Such contributions are related
to specific combinations of production and decay subprocesses and can
be additionally enhanced by the experimental cuts.  A typical example
is the production of fermion pairs originating from the decay of a
gauge boson, e.g.\ in the process
\bq
{\rm u} \quad \bar{\rm u} \quad\to\quad 
{\rm g} \quad {\rm g} \quad {\rm Z} \quad\to\quad
{\rm g} \quad {\rm g} \quad {\rm e}^+ \quad {\rm e}^-\,,
\eq
with the ${\rm Z}$ boson decaying into the ${\rm e}^+ {\rm e}^-$ pair.
\recola\ allows to select such specific contributions, even with
multiple and nested decays,
like for example
\bq
{\rm e}^+ \quad {\rm e}^- \quad\to\quad 
{\rm t} \;(\;\to\;{\rm W}^+ \;(\;\to\;{\rm u}\;\bar{\rm d}\;)\quad{\rm b} \;) \quad
\bar{\rm t} \;(\;\to\; {\rm e}^- \quad \bar{\nu}_e \quad \bar{\rm b} \;) .
\eq
This feature can be used to extract the resonant parts of the
amplitude. 
Moreover it can be employed to calculate matrix elements in the pole
approximation.  In this approximation, only the resonant parts of the
amplitude are kept and the residues of the poles in the amplitude are
calculated with on-shell momenta.  To this end, the complex squared
mass $\mu^2=m^2-{\rm i} m\Gamma$ of the resonant particle is
replaced by its real part ${\rm Re}\,(\mu^2)=m^2$ everywhere in the
amplitude except for the denominators of resonant propagators where
the width $\Gamma$ is kept.  The latter are further evaluated off
shell at $p^2\neq m^2$, while the rest of the amplitude is calculated
with on-shell kinematics, i.e.\ $p^2=m^2$.  This mismatch is accounted
for in \recola\ by the possibility of choosing a different value for
$p^2$ in the denominator of the resonant propagator than in the rest
of the amplitude.  At NLO, the selection of resonant contributions to
a process can be used to calculate the so-called factorizable
corrections in the pole approximation%
\footnote{The implementation of the factorizable NLO corrections in
  \recola{} in the pole approximation has been validated for Drell--Yan
  processes with resonant $\PZ$ and $\PW$~boson, as well as for doubly
  resonant processes with intermediate $\PZ$ and $\PW$~bosons.}.
In addition, \recola\ offers the possibility to switch off corrections
of self-energy type related to the resonant propagators, which cancel
in the
pole approximation.%
\footnote{While the omission of the self-energies related to the
  resonance improves the numerical stability in the pole
  approximation, these contributions have to be taken into account
 if the pole approximation is not used.}

In order to apply the pole approximation the following steps have to
be performed:
\begin{itemize}
\item The potentially resonant contributions have to be selected in
  the process definition (see \refse{definition} for details).
\item The potentially resonant particles have to be marked as resonant
 (see \refse{setrespart} for details).
\item The user has to ensure that the momenta $p$ of the resonant
  particles are on the mass shell, i.e.\ $p^2=m^2$.
\item The squared off-shell momenta in  the denominators of the resonant 
propagators have to be set  (see \refse{resonantmomentum} for details).
\end{itemize}
In this way the matrix elements in the pole approximation can be
obtained at LO and for the factorizable part at NLO. The
non-factorizable NLO corrections are not provided by the present
version of \recola.

Note that marking particles as resonant sets the widths of these
particles to zero in all defined processes. 
In order to calculate some of the matrix elements within the pole
approximation and others exactly, it is therefore necessary to
reset \recola\ and perform the integrations of the different
contributions sequentially.

\subsection{Conventions}
\label{conventions}

\recola\ uses the following symbols (of type {\tt character}) for the
SM particles:\\[1.5ex]
\quad\begin{tabular}{ll}
Higgs boson: &
\texttt{
'H'
}
\\
Goldstone bosons: &
\texttt{
'p0', 'p+', 'p-'
}
\\
Vector bosons: &
\texttt{
'g', 'A', 'Z', 'W+', 'W-'
}
\\
Neutrinos: &
\texttt{
'nu\_e', 'nu\_mu', 'nu\_tau'
}
\\
Anti-Neutrinos: &
\texttt{
'nu\_e\~{}', 'nu\_mu\~{}', 'nu\_tau\~{}'
}
\\
Charged leptons: &
\texttt{
'e-', 'e+',  'mu-', 'mu+', 'tau-', 'tau+'
}
\\
Quarks: &
\texttt{
'u', 'd', 'c', 's', 't', 'b'
}
\\
Anti-Quarks: &
\texttt{
'u\~{}', 'd\~{}', 'c\~{}', 's\~{}', 't\~{}', 'b\~{}'
}%
\; .
\end{tabular}
\\[1.5ex]
The helicities of the external particles are described in \recola\ by a 
variable of type {\tt character} with the values
\texttt{[+],[-],[0]}, or by an {\tt integer} variable with values
\texttt{+1,-1,0}.  This value determines the spinor (polarization vector)
to be attributed to the corresponding fermion (vector boson), 
and is fixed to zero for scalar particles:\\[1.5ex]
\quad\begin{tabular}{llll}
  Incoming fermion:      & \texttt{[+]} or \texttt{+1}&$\to$ &$u_+$ \\
  & \texttt{[-]} or \texttt{-1}&$\to$ &$u_-$ \\
  Outgoing fermion:      & \texttt{[+]} or \texttt{+1}&$\to$ &$\bar{u}_+$ \\
  & \texttt{[-]} or \texttt{-1}&$\to$ &$\bar{u}_-$ \\
  Incoming anti-fermion: & \texttt{[-]} or \texttt{-1}&$\to$ &$\bar{v}_-$ \\
  & \texttt{[+]} or \texttt{+1}&$\to$ &$\bar{v}_+$ \\
  Outgoing anti-fermion: & \texttt{[-]} or \texttt{-1}&$\to$ &$v_-$ \\
  & \texttt{[+]} or \texttt{+1}&$\to$ &$v_+$ \\
  \\[-.2cm]
  Transverse vector boson: & \texttt{[+]} or \texttt{+1} &$\to$&
  $\epsilon_+$ \\
  & \texttt{[-]} or \texttt{-1} &$\to$&
  $\epsilon_-$ \\
  Longitudinal vector boson: & \texttt{[0]} or \texttt{0} &$\to$&
  $\epsilon_0$ \\
  \\[-.2cm]
  Scalar boson: & \texttt{[0]} or \texttt{0} &$\to$&
  $1$ \; .
\end{tabular}\\[1.5ex]
\begin{sloppypar}
The explicit expressions of the spinors and polarization vectors used
in \recola\ can be found in \ref{appendix_spinors}.
\end{sloppypar}

\begin{sloppypar}
For the normalization of the amplitudes we use the conventions of
\citeres{Agashe:2014kda,Bohm:2001yx}. Accordingly,  cross sections and
decay widths are obtained as
\bq
\sigma(P_1,P_2 \to p_1,\dots,p_n) 
=
\frac{(2\pi)^4}{4\,\sqrt{(P_1\cdot P_2)^2 - P_1^2 P_2^2}}
\int\!\mathrm{d}\Phi_n(P_1 \!+\! P_2,p_1,\dots,p_n)\,\overline{{\cal A}^2},
\nonumber
\eq
\end{sloppypar}%

\bq
\Gamma(P \to p_1,\dots,p_n) 
=
\frac{(2\pi)^4}{2\sqrt{P^2}} 
\int\!\mathrm{d}\Phi_n(P,p_1,\dots,p_n)\,\overline{{\cal A}^2},
\eq
with
\bq
\mathrm{d}\Phi_n(p,p_1,\dots,p_n) = 
\delta^4\Big(p-\sum_{i=1}^np_i\Big)\,
\prod_{i=1}^n\frac{\mathrm{d}^3p_i}{2E_i(2\pi)^3} \:.
\eq
The SM Feynman rules implemented in \recola\ follow 
the conventions of~\citeres{Bohm:2001yx,Denner:1991kt}.


\section{Installation}
\label{installation}

Since \recola\ relies on the \collier\ library, the installation of
both packages is required to have a working amplitude generator.  The
following two options are available: \bei
\item
Download of the stand-alone \recola--\collier\ package for a combined
installation of both libraries.
\item Download of the \recola\ package alone which then has to be
  linked to a local \collier\ installation.
\eei
Both packages are available from the web site 
"http://recola.hepforge.org".

\subsection{The \recola--\collier\ package}
\label{recola-collier package}

The stand-alone package \texttt{recola-collier\_$X.Y.Z$} contains
the version $X.Y.Z$ of the \recola\ library together with a working copy of the
\collier\ library.
After downloading 
the file \texttt{recola-collier\_$X.Y.Z$.tar.gz}, extract the tarball in the 
current working directory with the shell command
\begin{verbatim}
   "tar -zxvf recola-collier_X.Y.Z.tar.gz" .
\end{verbatim}
This operation creates the directory
\texttt{recola-collier\_$X.Y.Z$}
containing the following files and folders:
\bei
\item\texttt{CMakeLists.txt}:\\
\textsc{CMake} makefile to produce the \collier\ and 
\recola\ libraries;
\item\texttt{build}:\\
build directory, where \textsc{CMake} puts all 
files necessary for the creation of the libraries;
\item\texttt{COLLIER\_A.B.C}:\\
main directory of the \collier\ package 
\texttt{COLLIER\_$A.B.C$};
\item\texttt{recola\_X.Y.Z}:\\
main directory of the \recola\ package
\texttt{recola\_$X.Y.Z$} (see \refse{recola package} for 
details).
\eei

The combined compilation of \collier\ and \recola\ proceeds by
changing to the \texttt{build} directory and executing there the shell
command "\texttt{cmake [options] ..}" (creating Makefiles for
\collier\ and \recola\ located in \texttt{COLLIER\_$A.B.C$/build} and
\texttt{recola\_$X.Y.Z$/build}, respectively), followed by
\texttt{make}:
\begin{verbatim}
   "cd recola-collier-X.Y.Z/build"
   "cmake [options] .."
   "make" .
\end{verbatim}
This requires \textsc{CMake} to be installed on the system.  If no
options are specified, \textsc{CMake} automatically searches for
installed {\sc Fortran} compilers and chooses a suited one, e.g.
\texttt{gfortran}.
The user can force \textsc{CMake} to
use a specific compiler by adding in the \texttt{cmake} command line
the option
\begin{verbatim}
   "-D CMAKE_Fortran_COMPILER=<comp>" ,
\end{verbatim}
where \texttt{<comp>} can be \texttt{gfortran, ifort, pgf95, ...} or the 
full path to a compiler.

By default, the installation sequence generates \collier\ and \recola\ 
as shared libraries: 
\bei
\item
\texttt{libcollier.so} is created in 
\texttt{recola-collier\_X.Y.Z/COLLIER\_A.B.C} and the corresponding
module files are placed in the \texttt{modules} subdirectory within this folder.
\item \texttt{librecola.so} is created in
  \texttt{recola-collier\_X.Y.Z/recola\_X.Y.Z} and the corresponding
  module files are placed in the \texttt{modules} subdirectory within
\sloppy
  this folder.
\eei
The option
\begin{verbatim}
   "-D static=ON"
\end{verbatim}
causes \textsc{CMake} 
to create the static libraries 
\texttt{libcollier.a} and \texttt{librecola.a} instead of the shared ones.

The packages \recola\ and \collier\ both contain a directory called
\texttt{demos}, where the user can find programs that illustrate the
usage of the codes.  Issuing
\begin{verbatim}
   "make <demofile>"
\end{verbatim}
in the directory \texttt{recola-collier-X.Y.Z/build}, with
\texttt{<demofile>} being the name (without extension) of one of these
demofiles, will compile the corresponding program. The executable
\texttt{<demofile>} is created in the respective directory
\texttt{demos} of \recola\ or \collier.  
More details on \recola\ demo programs are given in 
\refse{recola package}; for more details on the demo programs of
\collier\ we refer to \citere{Denner:2016kdg}.

\subsection{The \recola\ package}
\label{recola package}

\begin{sloppypar}
If the user wants to use his local installation%
\footnote{ \collier\ can be downloaded from
  \href{http://collier.hepforge.org/}{\mbox{http://collier.hepforge.org}}.}  
of \collier\ with \recola, he can
download the archive \texttt{recola\_X.Y.Z.tar.gz} containing only the
\recola\ package, in its version $X.Y.Z$.  Extract the tarball in the
current working directory with the shell command
\end{sloppypar} 
\begin{verbatim}
   "tar -zxvf recola_X.Y.Z.tar.gz" .
\end{verbatim}
This operation creates the directory \texttt{recola\_X.Y.Z}
containing the following files and folders:
\bei
\item \texttt{CMakeLists.txt}: \\
\textsc{CMake} makefile, to produce \recola\ library;
\item \texttt{build}: \\
build directory, where \textsc{CMake} puts all necessary
files for the creation of the library;
\item \texttt{src}: \\
\recola\ source directory, containing\\[-.7cm]
\bei
\item the source files
\texttt{input\_rcl.f90}, 
\texttt{process\_definition\_rcl.f90}, 
\texttt{process\_generation\_rcl.f90},
\texttt{process\_computation\_rcl.f90}, and \sloppy
\texttt{reset\_rcl.f90} with the global variables and public subroutines 
accessible to the user; 
\\[-.6cm]
\item the directory \texttt{internal} with private source files which should 
not be modified by the user.
\eei
\item \texttt{demos}: \\
directory with demo programs illustrating the use of \recola, including
shell scripts for their compilation and execution.
\eei 

The compilation of \recola\ proceeds by changing to the \texttt{build} 
directory and executing there the shell command 
\texttt{"cmake [options] .."} (creating a Makefile for \recola\ in 
\texttt{recola\_X.Y.Z/build}), followed by \texttt{make}:
\begin{verbatim}
   "cd recola_X.Y.Z/build"
   "cmake [options] .."
   "make [demofile]" .
\end{verbatim}
This requires \textsc{CMake} to be installed on the system. If no
options are specified, \textsc{CMake} automatically searches for
installed {\sc Fortran} compilers and chooses a suited one.
The user can force \textsc{CMake} to use a
specific compiler by adding in the \texttt{cmake} command line the
option
\begin{verbatim}
   "-D CMAKE_Fortran_COMPILER=<comp>" ,
\end{verbatim}
where \texttt{<comp>} can be \texttt{gfortran, ifort, pgf95, ...} or the 
full path to a compiler.

By default, the installation sequence generates \recola\ as shared
library \texttt{librecola.so} with the corresponding 
module files placed in the \texttt{modules} subdirectory. The option
\begin{verbatim}
   "-D static=ON"
\end{verbatim}
causes \textsc{CMake} to create the static library
\texttt{librecola.a} instead of the shared one.

The installation procedure further links \recola\ with the \collier\
library.  If not specified otherwise, \textsc{CMake} assumes the
existence of a folder named \texttt{COLLIER} that is located in the
parent directory of \texttt{recola\_X.Y.Z} and that contains the
\collier\ library \texttt{libcollier.so} and/or \texttt{libcollier.a},
as well as the subdirectory \texttt{modules} with the module files of
\collier.  Note that, depending on whether \recola\ shall be generated
as shared or as static library, the \collier\ library must be present
in the same format.

While the location of the \texttt{libcollier.so}/\texttt{libcollier.a}
and the module files within the \collier\ folder must be kept as
described above, the overall path may deviate from the default
setting. In this case, the full path to the \collier\ directory must
be given to \textsc{CMake} via the option
\begin{verbatim}
   "-D collier_path=<path to collier>" ,
\end{verbatim}
where \texttt{<path to collier>} can be either an absolute or a relative path. 

Moreover, by adding the option
\begin{verbatim}
   "-D cmake_collier=ON"
\end{verbatim}
to the \texttt{cmake} command line, the user can enforce that
\collier\ is \mbox{(re-)}com\-piled when the installation sequence for
\recola\ is performed\footnote{This only works if the complete
  \collier\ package with all source files is provided in the
  respective folder.}.  In this case the \textsc{CMake} makefile of
\recola\ calls the \textsc{CMake} makefile of \collier\ and generates
the \collier\ Makefile (any existing \collier\ Makefile is
overwritten).  The subsequent execution of \texttt{make} in
\texttt{recola-X.Y.Z/build} then generates the \collier\ library and
module files (placed in the respective directory) in addition to
\recola\ library and modules.

To create executables for the demo programs of \recola\ in the
directory {\tt demos}, the command
\begin{verbatim}
   "make <demofile>"
\end{verbatim}
should be issued in the directory \texttt{recola-X.Y.Z/build}, with
\texttt{<demofile>} being either \texttt{demo0\_rcl}, 
\texttt{demo1\_rcl}, \texttt{demo2\_rcl} or \texttt{demo3\_rcl}.
Alternatively, the user can execute (after issuing \texttt{cmake})
the shell scripts \texttt{run} with the command
\begin{verbatim}
   ./run <demofile>
\end{verbatim}
in the {\tt demos} directory.
This generates and runs the respective executable 
\texttt{<demofile>}. 

The demo programs \texttt{demo0\_rcl}, \texttt{demo1\_rcl}, 
\texttt{demo2\_rcl}, \texttt{demo3\_rcl} exemplify the usage 
of \recola\ for various purposes:
\bei
\item \texttt{demo0\_rcl}:\\
Basic usage of \recola.
\item \texttt{demo1\_rcl}:\\
  Usage of \recola\ for more than one process simultaneously, with
  explicit modification of input parameters and with selection of
  specific helicities for the external particles and of certain powers
  of the strong coupling constant.  In addition, files with \LaTeX\ 
  source code for diagrams are generated.
\item \texttt{demo2\_rcl}:\\
Usage of \recola\ for the selection of resonant contributions 
and pole approximation.
\item \texttt{demo3\_rcl}:\\
\sloppy
Usage of \recola\ for the computation of colour- and/or 
spin-correlation.
\eei

The {\tt demos} directory also contains the shell script 
\texttt{draw-tex} which compiles all \LaTeX\ files 
of the form \texttt{process\_*.tex} present in the folder 
and creates the corresponding \texttt{.pdf} files (see 
\refse{input draw} for more details). 
It can be run executing 
\begin{verbatim}
   ./draw-tex
\end{verbatim}
in the {\tt demos} directory.


\section{Usage of \recola}
\label{calling recola}
In order to use \recola\ in a {\sc Fortran} program, its modules
have to be loaded by including the line
\begin{verbatim}
  use recola
\end{verbatim}
in the preamble of the respective code, and the library {\tt
  librecola.so} or {\tt librecola.a} has to be supplied to the linker.
This gives access to the public functions and subroutines of the
\recola\ library described in the following subsections. The names of
all these routines end with the suffix ``{\tt \_rcl}''. This name
convention is supposed to avoid conflicts with routine names present
in the master program and increases readability by allowing for an
easy identification of command lines referring to the \recola\ 
library.

Typically, an application of \recola\ involves the following five steps:

\bei
\item{\bf Step 1: Setting input parameters (optional)}
  
The input needed for the computation of SM processes can be set by the
user in two ways: either by editing the file \texttt{input.f90},
changing there the values of the corresponding variables explicitly,
or by making use of subroutines provided by \recola\ for this purpose.
While the former option requires a recompilation of the program, the
latter allows for dynamical changes of the input parameters within the
same run of the program.  Input variables and subroutines are
described in \refse{input variables} and \refse{input subroutines},
respectively.  Since \recola\ provides default values for all input
parameters, this first step is optional.

\item{\bf Step 2: Defining the processes}

Before \recola\ can be employed to calculate matrix elements 
for one or more processes, each process must be declared and labelled with
a unique identifier. This is done by calling the 
\sloppy
subroutine \texttt{define\_process\_rcl} for every process, as described
in \refse{definition}.
\item{\bf Step 3: Generating the processes}
 
  In the next step the subroutine \texttt{generate\_processes\_rcl} is
  called which triggers the initialization of the complete list of
  processes defined in step 2.  As a result, all relevant building
  blocks for the recursive computation of off-shell currents are
  generated (see \refse{generation} for details).
\item{\bf Step 4: Computing the processes}
  
  After the arrangements made in the previous steps, \recola\ is ready
  to calculate amplitudes for any of the processes defined in step 2.
  The computation of the amplitude and of the squared amplitude is
  performed by means of the subroutine \texttt{compute\_process\_rcl},
  which uses the process-dependent information on the recursive
  procedure derived in step 3. The subroutine
  \texttt{compute\_process\_rcl} is called with the momenta of the
  external particles provided by the user.  In a Monte Carlo
  integration, the call of \texttt{compute\_process\_rcl} is repeated
  many times for different phase-space points.

{\sloppy 
  \recola\ further provides subroutines that allow to obtain
  particular contributions of the amplitude or the squared amplitude.
  In particular, it is possible to calculate colour- and/or
  spin-correlated squared amplitudes at the Born level.  Making use of
  the subroutines \texttt{set\_alphas\_rcl} or
  \texttt{compute\_running\_alphas\_rcl} one can also work with a
  running value for the strong coupling constant $\alpha_s$.}

Detailed information on the subroutines that can be employed in step 4
will be given in \refse{computation}.
\item{\bf Step 5: resetting \recola}

Finally, by calling the subroutine \texttt{reset\_recola\_rcl}, the
process-depen\-dent information generated in steps 2--4 is deleted and
the corresponding memory is deallocated. The input variables keep
their values defined in step 1 before.
\eei
Note that these steps have to be followed in the order given above. 
In particular, after step 3 no new
process can be defined unless \recola\ is reset (step 5). After
step 5 the user can restart with step 1 or step 2. 
More information on the allowed sequence of calls can be found in the
description of the routines below.

Examples of calls of \recola\ can be found in the directory 
\texttt{demos}. 


\subsection{Input variables}
\label{input variables}

The physical input parameters and the flags steering the output are
declared and initialized in the file \texttt{input.f90}.  This file
contains default values for these variables, which can be changed by
the user.

\subsubsection{Pole masses and widths of the SM particles}
\label{input masses}

The SM particles can be grouped into massive unstable particles,
characterized by their mass and decay width, massive stable particles,
characterized by their mass, and massless ones. In \recola, the gluon,
the photon and the neutrinos are treated as strictly massless (though
the photon and the gluon can actually get a fictitious mass to
regularize soft singularities, see \refse{input soft}). The electron
as well as the up, down and strange~quarks are considered in \recola\ 
as (potentially) massive stable particles, implying the possibility to
assign to them a non-zero mass. All other particles are considered as
(potentially) massive unstable particles, so that apart from a
non-zero mass they can also be assigned a non-zero width.
 
The mass and width variables declared in 
\texttt{input.f90} represent 
the pole mass $m_{_{\rm P}}$ and the pole width 
$\Gamma_{_{\rm P}}$, 
defined from 
the complex pole $s_{_{\rm P}}$ of the propagator as
\bq
s_{_{\rm P}} = m_{_{\rm P}}^2 - \mathrm{i}\,\Gamma_{_{\rm P}}\,m_{_{\rm P}}.
\eq
They are related to the on-shell quantities
$m_{_{\rm OS}}$ and $\Gamma_{_{\rm OS}}$ measured at the LEP and Tevatron experiments
for the $\PW$ and the $\PZ$ boson via
\bq
m_{_{\rm P}} = \frac{m_{_{\rm OS}}}{\sqrt{1+\Gamma_{_{\rm OS}}^2/m_{_{\rm OS}}^2}},
\qquad\qquad
\Gamma_{_{\rm P}} = 
\frac{\Gamma_{_{\rm OS}}}{\sqrt{1+\Gamma_{_{\rm OS}}^2/m_{_{\rm OS}}^2}}.
\eq
The default values (in GeV) for the pole masses and widths in \recola\ are:\small
\begin{verbatim}
  real(dp) :: mass_z  =  91.153480619182744d0
  real(dp) :: width_z  = 2.4942663787728243d0
  real(dp) :: mass_w  =  80.357973609877547d0
  real(dp) :: width_w  = 2.0842989982782196d0
  real(dp) :: mass_h  = 125.d0,                width_h  = 0.d0
  real(dp) :: mass_el =   0.d0
  real(dp) :: mass_mu =   0.d0,                width_mu = 0.d0
  real(dp) :: mass_ta =   0.d0,                width_ta = 0.d0
  real(dp) :: mass_u  =   0.d0
  real(dp) :: mass_d  =   0.d0
  real(dp) :: mass_c  =   0.d0,                width_c  = 0.d0
  real(dp) :: mass_s  =   0.d0
  real(dp) :: mass_t  = 173.2d0,               width_t  = 0.d0
  real(dp) :: mass_b  =   0.d0,                width_b  = 0.d0 ,
\end{verbatim}\normalsize
where $\tt dp$ indicates a double precision variable.  The default
values for the $\PZ$ and $\PW$ bosons have been computed from the
on-shell values
$$
\begin{array}{llll}
M_{\rm Z}^{^{\rm OS}} = & 91.1876 \, {\rm GeV}, \qquad & 
\Gamma_{\rm Z}^{^{\rm OS}} = & 2.4952 \, {\rm GeV},
\\
M_{\rm W}^{^{\rm OS}} = & 80.385  \, {\rm GeV}, \qquad & 
\Gamma_{\rm W}^{^{\rm OS}} = & 2.085 \, {\rm GeV}.
\end{array}
$$

\subsubsection{Parameters governing the treatment of collinear singularities}
\label{input collinear}

The treatment of collinear singularities caused by a fermion $f$ is
fixed from the value of its pole mass $m_f$: for $m_f=0$, dimensional
regularization is applied with the regularization parameters given in
\refse{input dimensional regularization}.
For $m_f\neq 0$, the variable \texttt{light\_}$\,f$ of type {\tt
  logical}, which can be set individually for each fermion $f$ =
\texttt{el}, \texttt{mu}, \texttt{ta}, \texttt{u}, \texttt{d},
\texttt{c}, \texttt{s}, \texttt{t}, \texttt{b}, controls how \recola\ 
deals with the fermion mass $m_f$:
\bei
\item \texttt{light\_}$\,f$ = .true.:\\
The fermion $f$ is considered as ``light'', and its non-zero pole mass
$m_f$ is kept only in mass-singular logarithms but neglected
elsewhere. In this case, $m_f$ is used as mass regulator.
\item \texttt{light\_}$\,f$ = .false.:\\
The fermion $f$ is considered as heavy, and
the full dependence on the pole mass $m_f$ is kept.
\eei
If the pole mass of fermion $f$ is set to $m_f=0$, the actual value of
\texttt{light\_}$\,f$ is irrelevant.  The default values of
\texttt{light\_}$\,f$ are
\begin{verbatim}
  logical :: light_el = .true.
  logical :: light_mu = .true.
  logical :: light_ta = .true.
  logical :: light_u  = .true.
  logical :: light_d  = .true.
  logical :: light_c  = .true.
  logical :: light_s  = .true.
  logical :: light_t  = .false.
  logical :: light_b  = .true. .
\end{verbatim}

\subsubsection{Parameters governing the treatment of soft singularities}
\label{input soft}

Soft singularities can be cured in \recola\ in two different ways, 
and the method to be applied is selected depending on
the value of the {\tt integer} variable \texttt{reg\_soft}:
\bei
\item \texttt{reg\_soft} = 1: \\
  Dimensional regularization is used with the regularization parameters 
given in \refse{input dimensional regularization}.
\item \texttt{reg\_soft} = 2: \\
  Mass regularization is used with regulator \texttt{lambda} for the
  photon/gluon mass.  \eei In the case of dimensional regularization
  (\texttt{reg\_soft} = 1), the value of the variable \texttt{lambda}
  is irrelevant. Note further that mass regularization for soft
  singularities (\texttt{reg\_soft} = 2) is only allowed in
  combination with mass regularization for all collinear
  singularities. Therefore, in the case of \texttt{reg\_soft} = 2, the
  masses of external charged fermions must be chosen different from
  zero for all processes.
The default values for \texttt{reg\_soft} and \texttt{lambda} (in GeV) are 
\begin{verbatim}
  integer :: reg_soft = 1;   real(dp) :: lambda = 100d0 .
\end{verbatim}

\subsubsection{Dimensional regularization parameters}
\label{input dimensional regularization}

The conventions used by \recola\ for dimensional regularization have been introduced in
\refse{dimensional regularization}. The regularization parameters are represented 
by the variables (of type {\tt real(dp)})
\bq
\deltauv = \Delta_\uv,
\qquad
\muuv = \mu_\uv,
\eq
\bq
\deltair\  = \Delta_\ir,
\qquad
\deltairr\ = \Delta_\irr,
\qquad
\muir\ =\mu_\ir,
\eq
with the default values (in GeV for \muuv\ and \muir) 
\begin{verbatim}
  real(dp) :: DeltaUV = 0d0,                 muUV = 100d0
  real(dp) :: DeltaIR = 0d0, DeltaIR2 = 0d0, muIR = 100d0 .
\end{verbatim}

\subsubsection{Parameters for the renormalization of the QCD coupling}
\label{input alphas}

\recola\ uses an $\overline{\rm MS}$ prescription for the
renormalization of the strong coupling constant (see
\refse{dimensional regularization} for details).  The renormalization
scheme is fixed by the choice of the scale $Q$, given by the {\tt
real(dp)} variable \Qren, and the selection of the number of active
flavours $N_{\rm f}$, given by the {\tt integer} variable \Nfren\
accepting the following values: \bei
\item
  \Nfren\ = $-1$: \\
  The {\it variable-flavour scheme} is used, and all quarks with
  masses lower than \Qren\ are considered active flavours.
\item
  \Nfren\ = $3,4,5,6$: \\
  A {\it fixed-flavour scheme} is used, and the $N_{\rm f}=\texttt{Nfren}$ lightest
  quarks (those with the smallest mass values) are treated as active
  flavours. Note that the other quarks, which are not taken as active
  flavours, must have a non-vanishing mass.
\eei
As numerical input, \recola\ needs the value for the {\tt real(dp)}
variable \als\ corresponding the QCD coupling $\alpha_s(Q^2)$ at the
scale $Q$ in the selected flavour scheme (see also \refse{alphas}).
The default values for the variables introduced in this section are
(in GeV for \Qren) 
\begin{verbatim}
  real(dp) :: als   = 0.118d0
  real(dp) :: Qren  = 91.1876d0
  integer  :: Nfren = 5 .
\end{verbatim}

\subsubsection{Parameters for the renormalization of the EW coupling}
\label{input ew renormalization}

The {\tt integer} variable \texttt{ew\_reno\_scheme} allows to switch 
between the three renormalization schemes for the EW coupling $\alpha$ 
implemented in \recola\ (see \refse{ew renormalization}):
\bei
\item \texttt{ew\_reno\_scheme} = 1:\\
  The $G_{\rm F}$ scheme is chosen, and $\alpha$ is determined from
  the Fermi constant $G_{\rm F}$, given by the value of the variable
  \gf\ of type {\tt real(dp)} .

\item \texttt{ew\_reno\_scheme} = 2:\\
The $\alpha(0)$ scheme is chosen, and $\alpha$ is set to the value
$\alpha(0)$ stored in the variable \alO\ of type {\tt real(dp)} .

\item \texttt{ew\_reno\_scheme} = 3:\\
The $\alpha(M_{\PZ})$ scheme is chosen, and $\alpha$ is set to the value 
$\alpha(M_{\rm Z})$ stored in the variable \alZ\ of type {\tt real(dp)} .
\eei
Depending on the selected scheme, the EW coupling $\alpha$ is
determined from the corresponding variable \gf, \alO\ or \alZ, while
the other two variables do not enter the calculation and their actual
values are thus irrelevant.  The default values (in ${\rm GeV}^{-2}$
for \gf) are
\begin{verbatim}
  integer  :: ew_reno_scheme = 1
  real(dp) :: gf  = 1.16637d-5
  real(dp) :: al0 = 1d0/137.035999679d0
  real(dp) :: alZ = 1d0/128.936d0 .
\end{verbatim}

\subsubsection{Parameter for the renormalization of masses of unstable particles}
\label{input cms}
For the renormalization of the masses of unstable particles, the {\tt
  integer} variable \texttt{complex\_mass\_scheme} allows to choose
between the complex-mass or the on-shell renormalization scheme (see
\refse{ew renormalization}):
\bei
\item
\texttt{complex\_mass\_scheme} = 1: \\
The complex-mass scheme 
of~\citeres{Denner:1999gp,Denner:2005fg,Denner:2006ic} is used.
\item
\texttt{complex\_mass\_scheme} = 0: \\
The on-shell scheme is used. 
\eei
By default, the complex-mass scheme is selected:
\begin{verbatim}
  integer :: complex_mass_scheme = 1 .
\end{verbatim}

\subsubsection{Parameter for self-energies of resonant particles}
\label{input resSE}

Whether self-energy insertions (including the corresponding
\sloppy
coun\-ter\-terms) are taken into account for resonant particles, is 
determined by the variable \texttt{resSE} (of type {\tt logical}):
\bei
\item
\texttt{resSE = .true.}: \\
Self-energies of resonant particles are included in the computation 
of the NLO amplitude.
\item
\texttt{resSE = .false.}: \\
Self-energies of resonant particles are excluded in the computation 
of the NLO amplitude.
\eei
By default, self-energies of resonant particles are included:
\begin{verbatim}
  logical :: resSE = .true. .
\end{verbatim}
While the self-energies should always be included if the resonant
particles are off shell, their omission improves the numerical
stability for on-shell resonances, where the self-energies should
cancel exactly.

\subsubsection{Options for the visualization of the off-shell currents}
\label{input draw}

The off-shell currents and branches used by {\tt recola} as building
blocks for the construction of the amplitude (see
\citere{Actis:2012qn} for details) 
can be visualized as diagrams,
similar to the Feynman-diagrammatic visualization of the amplitude in
the traditional approach.
The {\tt integer} variable \texttt{draw} determines the options for
the generation of a file with \LaTeX\ source code for the
corresponding diagrams:
\bei
\item \texttt{draw = 0}:\\
No \LaTeX\ file is generated.

\item \texttt{draw = 1}:\\
For each process, a \LaTeX\ file \texttt{process\_}$n$\texttt{.tex} is
created, where $n$ is the identifier assigned to the process in the
call of \texttt{define\_process\_rcl} (see \refse{definition}).

After a legend, the \LaTeX\ file states the process 
under consideration. According to the internal conventions of \recola,
outgoing particles are crossed into the initial state, and 
each particle carries an identifier in form of a binary number. For instance,
for the process $\Pu\Pubar\to\PZ\Pg\Pg$ the output reads:
\begin{center}
\begin{tabular}{*{10}{c}}
\hline\hline \\[-.2cm]
\quad
${\scriptstyle{\rm Z}}$
& $+$ &
${\rm u}$
& $+$ &
$\bar{\rm u}$
& $+$ &
${\rm g}$
& $+$ &
${\rm g}$
& $\,\to\quad 0 \quad$\\
\quad
  1
& &
  2
& &
  4
& &
  8
& &
 16
&
\\[+.2cm]\hline\hline
\end{tabular}
\end{center}

Then all branches needed for the tree-level amplitude are drawn, 
listed following the order in which they are computed: \\[2ex]
\setlength{\fboxsep}{1ex}
\fbox{Tree level}\\
{\setlength{\unitlength}{.95pt}\SetScale{0.95}
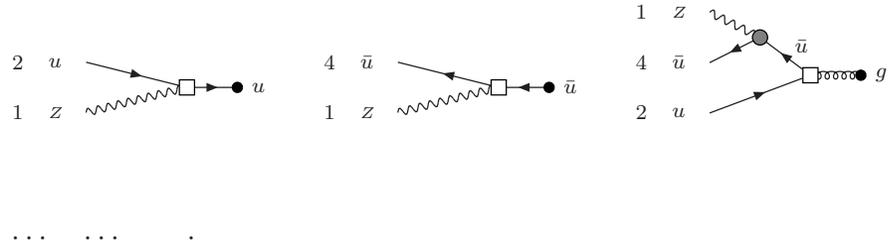
\begin{picture}(120,  45)(-60, -40)
\Text(-45,   0)[rc]{\scriptsize   2}\Text(-35,   0)[lc]{\scriptsize $u$}
\ArrowLine( -20,   0)(20, -10)         
\Text(-45, -20)[rc]{\scriptsize   1}\Text(-35, -20)[lc]{\scriptsize ${\scriptstyle{Z}}$}
\Photon( -20, -20)(20, -10){1.5}{10}
\ArrowLine(20, -10)(40, -10)         
\Text(46, -10)[lc]{\scriptsize $u$}
\GBoxc(20, -10)(6,6){1}
\SetWidth{1.5}
\GCirc(40, -10){1.5}{0}
\end{picture}
\begin{picture}(120,  45)(-60, -40)
\Text(-45,   0)[rc]{\scriptsize   4}\Text(-35,   0)[lc]{\scriptsize $\bar{u}$}
\ArrowLine(20, -10)( -20,   0)         
\Text(-45, -20)[rc]{\scriptsize   1}\Text(-35, -20)[lc]{\scriptsize ${\scriptstyle{Z}}$}
\Photon( -20, -20)(20, -10){1.5}{10}
\ArrowLine(40, -10)(20, -10)         
\Text(46, -10)[lc]{\scriptsize $\bar{u}$}
\GBoxc(20, -10)(6,6){1}
\SetWidth{1.5}
\GCirc(40, -10){1.5}{0}
\end{picture}
\begin{picture}(120,  65)(-60, -60)
\Text(-45,   0)[rc]{\scriptsize   1}\Text(-35,   0)[lc]{\scriptsize ${\scriptstyle{Z}}$}
\Photon(-20,   0)(0, -10){1.5}{ 5}
\Text(-45, -20)[rc]{\scriptsize   4}\Text(-35, -20)[lc]{\scriptsize $\bar{u}$}
\ArrowLine(0, -10)(-20, -20)         
\ArrowLine(20, -25)(0, -10)         
\Text(  14, -16)[lb]{\scriptsize $\bar{u}$}
\GCirc(0, -10){3}{.5}
\Text(-45, -40)[rc]{\scriptsize   2}\Text(-35, -40)[lc]{\scriptsize $u$}
\ArrowLine( -20, -40)(20, -25)         
\Gluon(20, -25)(40, -25){1.5}{5} 
\Text(46, -25)[lc]{\scriptsize $g$}
\GBoxc(20, -25)(6,6){1}
\SetWidth{1.5}
\GCirc(40, -25){1.5}{0}
\end{picture}
\\[-1ex]
\begin{picture}(20,  65)(-10, -30)
\ldots\quad\ldots\qquad.
\end{picture}%
}%

Next, the branches needed for the 4-dimensional bare one-loop
contri\-bu\-tion are drawn, grouped 
according to the respective 
cut-particle and listed following the order in which they are computed: \\[2ex]
\setlength{\fboxsep}{1ex}
\fbox{Bare loop; \qquad cut-particle: ${\scriptstyle{Y}_g}$}\\
{\setlength{\unitlength}{.95pt}\SetScale{0.95}
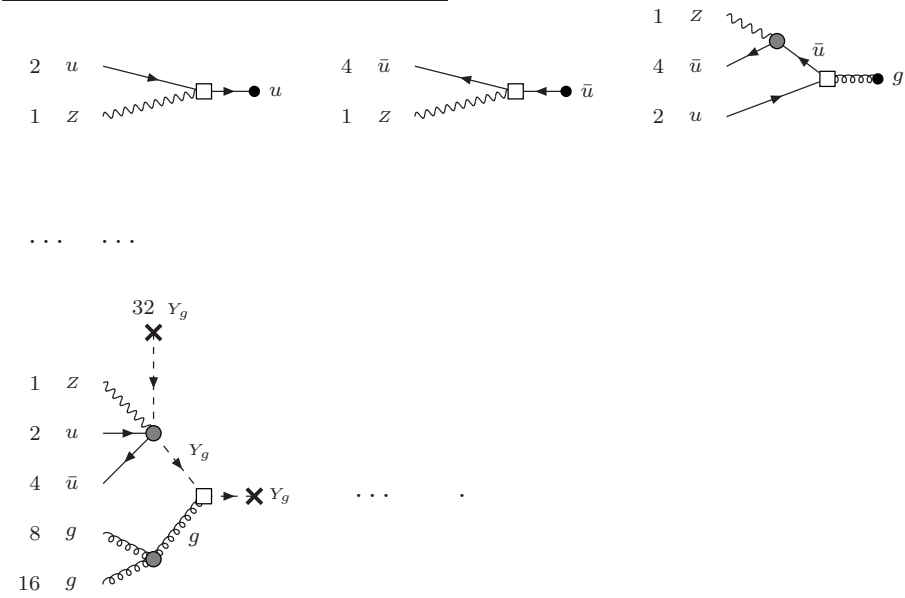
\begin{picture}(120,  45)(-60, -40)
\Text(-45,   0)[rc]{\scriptsize   2}\Text(-35,   0)[lc]{\scriptsize $u$}
\ArrowLine( -20,   0)(20, -10)         
\Text(-45, -20)[rc]{\scriptsize   1}\Text(-35, -20)[lc]{\scriptsize ${\scriptstyle{Z}}$}
\Photon( -20, -20)(20, -10){1.5}{10}
\ArrowLine(20, -10)(40, -10)         
\Text(46, -10)[lc]{\scriptsize $u$}
\GBoxc(20, -10)(6,6){1}
\SetWidth{1.5}
\GCirc(40, -10){1.5}{0}
\end{picture}
\begin{picture}(120,  45)(-60, -40)
\Text(-45,   0)[rc]{\scriptsize   4}\Text(-35,   0)[lc]{\scriptsize $\bar{u}$}
\ArrowLine(20, -10)( -20,   0)         
\Text(-45, -20)[rc]{\scriptsize   1}\Text(-35, -20)[lc]{\scriptsize ${\scriptstyle{Z}}$}
\Photon( -20, -20)(20, -10){1.5}{10}
\ArrowLine(40, -10)(20, -10)         
\Text(46, -10)[lc]{\scriptsize $\bar{u}$}
\GBoxc(20, -10)(6,6){1}
\SetWidth{1.5}
\GCirc(40, -10){1.5}{0}
\end{picture}
\begin{picture}(120,  65)(-60, -60)
\Text(-45,   0)[rc]{\scriptsize   1}\Text(-35,   0)[lc]{\scriptsize ${\scriptstyle{Z}}$}
\Photon(-20,   0)(0, -10){1.5}{ 5}
\Text(-45, -20)[rc]{\scriptsize   4}\Text(-35, -20)[lc]{\scriptsize $\bar{u}$}
\ArrowLine(0, -10)(-20, -20)         
\ArrowLine(20, -25)(0, -10)         
\Text(  14, -16)[lb]{\scriptsize $\bar{u}$}
\GCirc(0, -10){3}{.5}
\Text(-45, -40)[rc]{\scriptsize   2}\Text(-35, -40)[lc]{\scriptsize $u$}
\ArrowLine( -20, -40)(20, -25)         
\Gluon(20, -25)(40, -25){1.5}{5} 
\Text(46, -25)[lc]{\scriptsize $g$}
\GBoxc(20, -25)(6,6){1}
\SetWidth{1.5}
\GCirc(40, -25){1.5}{0}
\end{picture}
\\[-1ex]
\begin{picture}(120,  65)(-10, -30)
\ldots\quad\ldots
\end{picture}
\\[-1ex]
\begin{picture}(120, 130)(-60,-120)
\Text(3,9)[cc]{\scriptsize $  32\;\;{\scriptstyle{Y}_g}$}
\DashArrowLine(0,0)(0, -40){3}      
\Text(-45, -20)[rc]{\scriptsize   1}\Text(-35, -20)[lc]{\scriptsize ${\scriptstyle{Z}}$}
\Photon(-20, -20)(0, -40){1.5}{ 7}
\Text(-45, -40)[rc]{\scriptsize   2}\Text(-35, -40)[lc]{\scriptsize $u$}
\ArrowLine(-20, -40)(0, -40)         
\Text(-45, -60)[rc]{\scriptsize   4}\Text(-35, -60)[lc]{\scriptsize $\bar{u}$}
\ArrowLine(0, -40)(-20, -60)         
\DashArrowLine(0, -40)(20, -65){3}      
\Text(  14, -51)[lb]{\scriptsize ${\scriptstyle{Y}_g}$}
\GCirc(0, -40){3}{.5}
\Text(-45, -80)[rc]{\scriptsize   8}\Text(-35, -80)[lc]{\scriptsize $g$}
\Gluon(-20, -80)(0, -90){1.5}{ 5}
\Text(-45,-100)[rc]{\scriptsize  16}\Text(-35,-100)[lc]{\scriptsize $g$}
\Gluon(-20,-100)(0, -90){1.5}{ 5}
\Gluon(0, -90)(20, -65){1.5}{ 8}
\Text(  14, -85)[lb]{\scriptsize $g$}
\GCirc(0, -90){3}{.5}
\DashArrowLine(20, -65)(40, -65){3}      
\Text(46, -65)[lc]{\scriptsize ${\scriptstyle{Y}_g}$}
\GBoxc(20, -65)(6,6){1}
\SetWidth{1.5}
\Line(-3,-3)(3,3)\Line(-3,3)(3,-3)
\Line(37, -68)(43, -62)\Line(37, -62)(43, -68)
\Text(80,-65)[lc]{\ldots\qquad.}
\end{picture}%
}

Finally, the branches for the counterterm contribution and for 
the contribution of the rational parts $R_2$ are drawn, 
again listed following the order in which they are computed: \\[2ex] 
\setlength{\fboxsep}{1ex}
\fbox{Counterterms}\\[2ex]
{\setlength{\unitlength}{.95pt}\SetScale{0.95}
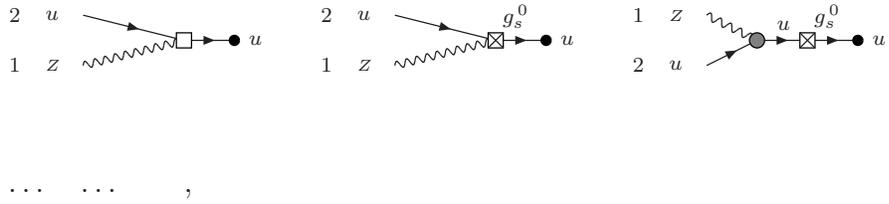
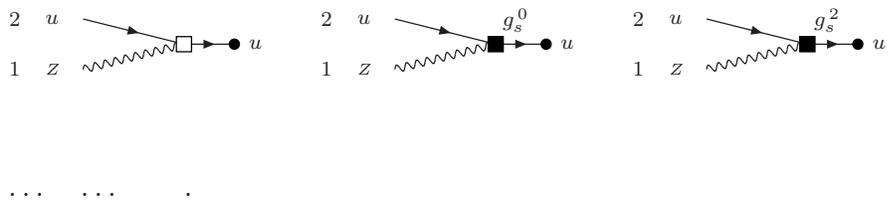
\begin{picture}(120,  45)(-60, -40)
\Text(-45,   0)[rc]{\scriptsize   2}\Text(-35,   0)[lc]{\scriptsize $u$}
\ArrowLine( -20,   0)(20, -10)         
\Text(-45, -20)[rc]{\scriptsize   1}\Text(-35, -20)[lc]{\scriptsize ${\scriptstyle{Z}}$}
\Photon( -20, -20)(20, -10){1.5}{10}
\ArrowLine(20, -10)(40, -10)         
\Text(46, -10)[lc]{\scriptsize $u$}
\GBoxc(20, -10)(6,6){1}
\SetWidth{1.5}
\GCirc(40, -10){1.5}{0}
\end{picture}
\begin{picture}(120,  45)(-60, -40)
\Text(-45,   0)[rc]{\scriptsize   2}\Text(-35,   0)[lc]{\scriptsize $u$}
\ArrowLine( -20,   0)(20, -10)         
\Text(-45, -20)[rc]{\scriptsize   1}\Text(-35, -20)[lc]{\scriptsize ${\scriptstyle{Z}}$}
\Photon( -20, -20)(20, -10){1.5}{10}
\ArrowLine(20, -10)(40, -10)         
\Text(46, -10)[lc]{\scriptsize $u$}
\GBoxc(20, -10)(6,6){1}
\Line(17, -13)(23,  -7)\Line(17,  -7)(23, -13)
\Text(28,  -5)[cb]{\scriptsize $g_s^{\,0}$}
\SetWidth{1.5}
\GCirc(40, -10){1.5}{0}
\end{picture}
\begin{picture}(120,  45)(-60, -40)
\Text(-45,   0)[rc]{\scriptsize   1}\Text(-35,   0)[lc]{\scriptsize ${\scriptstyle{Z}}$}
\Photon(-20,   0)(0, -10){1.5}{ 5}
\Text(-45, -20)[rc]{\scriptsize   2}\Text(-35, -20)[lc]{\scriptsize $u$}
\ArrowLine(-20, -20)(0, -10)         
\ArrowLine(0, -10)(20, -10)         
\Text(   8,  -6)[lb]{\scriptsize $u$}
\GCirc(0, -10){3}{.5}
\ArrowLine(20, -10)(40, -10)         
\Text(46, -10)[lc]{\scriptsize $u$}
\GBoxc(20, -10)(6,6){1}
\Line(17, -13)(23,  -7)\Line(17,  -7)(23, -13)
\Text(28,  -5)[cb]{\scriptsize $g_s^{\,0}$}
\SetWidth{1.5}
\GCirc(40, -10){1.5}{0}
\end{picture}
\\[-1ex]
\begin{picture}(120,  65)(-10, -30)
\ldots\quad\ldots\qquad,
\end{picture}%
}%

\setlength{\fboxsep}{1ex}
\fbox{Rational terms}\\*[2ex]
{\setlength{\unitlength}{.95pt}\SetScale{0.95}
\begin{picture}(120,  45)(-60, -40)
\Text(-45,   0)[rc]{\scriptsize   2}\Text(-35,   0)[lc]{\scriptsize $u$}
\ArrowLine( -20,   0)(20, -10)         
\Text(-45, -20)[rc]{\scriptsize   1}\Text(-35, -20)[lc]{\scriptsize ${\scriptstyle{Z}}$}
\Photon( -20, -20)(20, -10){1.5}{10}
\ArrowLine(20, -10)(40, -10)         
\Text(46, -10)[lc]{\scriptsize $u$}
\GBoxc(20, -10)(6,6){1}
\SetWidth{1.5}
\GCirc(40, -10){1.5}{0}
\end{picture}
\begin{picture}(120,  45)(-60, -40)
\Text(-45,   0)[rc]{\scriptsize   2}\Text(-35,   0)[lc]{\scriptsize $u$}
\ArrowLine( -20,   0)(20, -10)         
\Text(-45, -20)[rc]{\scriptsize   1}\Text(-35, -20)[lc]{\scriptsize ${\scriptstyle{Z}}$}
\Photon( -20, -20)(20, -10){1.5}{10}
\ArrowLine(20, -10)(40, -10)         
\Text(46, -10)[lc]{\scriptsize $u$}
\GBoxc(20, -10)(6,6){0}
\Text(28,  -5)[cb]{\scriptsize $g_s^{\,0}$}
\SetWidth{1.5}
\GCirc(40, -10){1.5}{0}
\end{picture}
\begin{picture}(120,  45)(-60, -40)
\Text(-45,   0)[rc]{\scriptsize   2}\Text(-35,   0)[lc]{\scriptsize $u$}
\ArrowLine( -20,   0)(20, -10)         
\Text(-45, -20)[rc]{\scriptsize   1}\Text(-35, -20)[lc]{\scriptsize ${\scriptstyle{Z}}$}
\Photon( -20, -20)(20, -10){1.5}{10}
\ArrowLine(20, -10)(40, -10)         
\Text(46, -10)[lc]{\scriptsize $u$}
\GBoxc(20, -10)(6,6){0}
\Text(28,  -5)[cb]{\scriptsize $g_s^{\,2}$}
\SetWidth{1.5}
\GCirc(40, -10){1.5}{0}
\end{picture}
\\[-1ex]
\begin{picture}(120,  65)(-10, -30)
\ldots\quad\ldots\qquad.
\end{picture}%
}%
\item \texttt{draw = 2}:\\
  For each process, a \LaTeX\ file is generated with the content as in
  the case \texttt{draw = 1}, and in addition the colour structures of
  the incoming and outgoing currents are explicitly written below each
  branch:
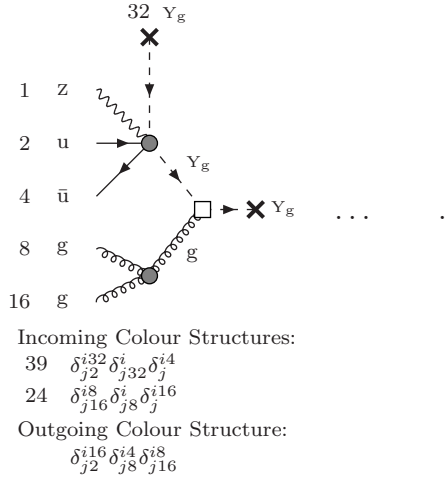
\begin{picture}(120, 190)(-60,-170)
\Text(3,9)[cc]{\scriptsize $  32\;\;{\scriptstyle{\rm Y}_{\rm g}}$}
\DashArrowLine(0,0)(0, -40){3}      
\Text(-45, -20)[rc]{\scriptsize   1}\Text(-35, -20)[lc]{\scriptsize ${\scriptstyle{\rm Z}}$}
\Photon(-20, -20)(0, -40){1.5}{ 7}
\Text(-45, -40)[rc]{\scriptsize   2}\Text(-35, -40)[lc]{\scriptsize ${\rm u}$}
\ArrowLine(-20, -40)(0, -40)         
\Text(-45, -60)[rc]{\scriptsize   4}\Text(-35, -60)[lc]{\scriptsize $\bar{\rm u}$}
\ArrowLine(0, -40)(-20, -60)         
\DashArrowLine(0, -40)(20, -65){3}      
\Text(  14, -51)[lb]{\scriptsize ${\scriptstyle{\rm Y}_{\rm g}}$}
\GCirc(0, -40){3}{.5}
\Text(-45, -80)[rc]{\scriptsize   8}\Text(-35, -80)[lc]{\scriptsize ${\rm g}$}
\Gluon(-20, -80)(0, -90){1.5}{ 5}
\Text(-45,-100)[rc]{\scriptsize  16}\Text(-35,-100)[lc]{\scriptsize ${\rm g}$}
\Gluon(-20,-100)(0, -90){1.5}{ 5}
\Gluon(0, -90)(20, -65){1.5}{ 8}
\Text(  14, -85)[lb]{\scriptsize ${\rm g}$}
\GCirc(0, -90){3}{.5}
\DashArrowLine(20, -65)(40, -65){3}      
\Text(46, -65)[lc]{\scriptsize ${\scriptstyle{\rm Y}_{\rm g}}$}
\GBoxc(20, -65)(6,6){1}
\Text(-50,-114)[l]{\scriptsize Incoming Colour Structures:}
\Text(-47,-126)[lb]{\scriptsize $  39$}
\Text(-30,-125)[l]{\scriptsize $\delta_{j2}^{i32}\delta_{j32}^i\delta_j^{i4}$}
\Text(-47,-138)[lb]{\scriptsize $  24$}
\Text(-30,-137)[l]{\scriptsize $\delta_{j16}^{i8}\delta_{j8}^i\delta_j^{i16}$}
\Text(-50,-149)[l]{\scriptsize Outgoing Colour Structure:}
\Text(-30,-160)[l]{\scriptsize $\delta_{j2}^{i16}\delta_{j8}^{i4}\delta_{j16}^{i8}$}
\SetWidth{1.5}
\Line(-3,-3)(3,3)\Line(-3,3)(3,-3)
\Line(37, -68)(43, -62)\Line(37, -62)(43, -68)
\Text(70, -68)[lc]{\ldots\qquad.}
\end{picture}
\eei
The default value for the variable \texttt{draw} is
\begin{verbatim}
  integer :: draw = 0 .
\end{verbatim}

\subsubsection{Output options}
\label{input print}
The \texttt{integer} variables \texttt{writeMat}, \texttt{writeMat2},
\texttt{writeCor}, and \texttt{writeRAM} define to which extent output
is written by \recola\ into the standard output channel.
Independently of the values of these parameters, \recola\ always
prints an initialization message followed by the values of the input
parameters.  For default values, this output reads:

{\footnotesize
\begin{verbatim}
xxxxxxxxxxxxxxxxxxxxxxxxxxxxxxxxxxxxxxxxxxxxxxxxxxxxxxxxxxxxxxxxxxxxxxxxxx
                          _    _   _   _        _ 
                         | )  |_  |   | |  |   |_|
                         | \  |_  |_  |_|  |_  | |

               REcursive Computation of One Loop Amplitudes

                                Version 1.0

      by S.Actis, A.Denner, L.Hofer, J.-N.Lang, A.Scharf, S.Uccirati

xxxxxxxxxxxxxxxxxxxxxxxxxxxxxxxxxxxxxxxxxxxxxxxxxxxxxxxxxxxxxxxxxxxxxxxxxx

--------------------------------------------------------------------------
 Pole masses and widths [GeV]:
 M_Z   =  91.153480619183                Width_Z   =  2.4942663787728    
 M_W   =  80.357973609878                Width_W   =  2.0842989982782    
 M_H   =  125.00000000000                Width_H   =  0.0000000000000    
 m_e   =  0.0000000000000           
 m_mu  =  0.0000000000000                Width_mu  =  0.0000000000000    
 m_tau =  0.0000000000000                Width_tau =  0.0000000000000    
 m_u   =  0.0000000000000           
 m_d   =  0.0000000000000           
 m_c   =  0.0000000000000                Width_c   =  0.0000000000000    
 m_s   =  0.0000000000000           
 m_t   =  173.20000000000                Width_t   =  0.0000000000000    
 m_b   =  0.0000000000000                Width_b   =  0.0000000000000    
--------------------------------------------------------------------------
 Renormalization done in the complex-mass scheme
--------------------------------------------------------------------------
  EW Renormalization Scheme: gfermi       Gf = 0.11663700000000E-04 GeV^-2
                                          alpha_Gf = 0.75552541674271E-02
--------------------------------------------------------------------------
 alpha_s Renormalization Scheme: 5-flavours Scheme
 alpha_s(Q) = 0.11800000000000           Q =  91.187600000000     GeV
--------------------------------------------------------------------------
 Delta_UV   =  0.0000000000000           mu_UV =  100.00000000000     GeV
 Delta_IR^2 =  0.0000000000000    
 Delta_IR   =  0.0000000000000           mu_IR =  100.00000000000     GeV
--------------------------------------------------------------------------
 Dimensional regularization for soft singularities
--------------------------------------------------------------------------
\end{verbatim}
}
If \texttt{writeMat}, \texttt{writeMat2}, or \texttt{writeCor} are 
different from \texttt{0}, the process under consideration and the 
momenta of the incoming and outgoing particles are printed 
whenever a routine is called which computes amplitudes
(see \refse{conventions} for particle conventions).
For example, for a phase-space point for the process 
$\mathrm{u \bar{u}} \to \mathrm{Z g g}$ the output reads:

{\footnotesize
\begin{verbatim}
 u u~ -> Z g g

 p1 = (4000.000000000,    0.000000000,    0.000000000, 4000.000000000) GeV
 p2 = (4000.000000000,    0.000000000,    0.000000000,-4000.000000000) GeV
 p3 = (2489.514720448,-2307.817376695,  306.489328734, -877.164509143) GeV
 p4 = (2694.735905719, 1902.356873535, 1290.340022324,-1406.293907431) GeV
 p5 = (2815.749373833,  405.460503160,-1596.829351058, 2283.458416573) GeV . 
\end{verbatim}
}

\subsubsection*{Output options for the amplitude}

The value of the variable \texttt{writeMat} determines the level of detail 
at which the results for 
the structure-dressed amplitudes ${\cal A}^{(\vec{c},\vec{h})}$ (see 
\refse{amplitude structure}) are printed for all computed processes: 

\bei

\item \texttt{writeMat = 0}:\\
Nothing is printed.

\item \texttt{writeMat = 1}:\\
  The results of the non-vanishing ${\cal A}^{(\vec{c},\vec{h})}$ are
  listed for each helicity configuration $\vec{h}$ and colour
  structure $\vec{c}$.
  
  In order to illustrate the notation employed for helicity
  configurations and colour structures, we give the output for the
  helicity configuration $\vec{h} = (+,-,+,+,+)$ and the colour
  structure
  $\delta^{i_2}_{j_4}\,\delta^{i_4}_{j_5}\,\delta^{i_5}_{j_1}$ (see
  \refse{conventions} for helicity conventions) in the sample process
  $\Pu\Pubar\to\PZ\Pg\Pg$: 

{\footnotesize
\begin{verbatim}
  AMPLITUDE [GeV^-1]

  Helicity configuration:   u[+] u~[-] -> Z[+] g[+] g[+]

                       i2    i4    i5 
  Colour structure:   d     d     d     .  
                       j4    j5    j1 
\end{verbatim}
}

For each helicity configuration and colour 
structure, the results for the Born and, if computed, one-loop
amplitude are written in separate tables:

{\footnotesize
\begin{verbatim}
   gs |               Born Amplitude A0               
  ----------------------------------------------------
    0 | ( 0.00000000000000E+00, 0.00000000000000E+00)
    1 | ( 0.00000000000000E+00, 0.00000000000000E+00)
    2 | (-0.88324465151595E-08,-0.29277610069124E-07)
    3 | ( 0.00000000000000E+00, 0.00000000000000E+00)
  ----------------------------------------------------
  SUM | (-0.88324465151595E-08,-0.29277610069124E-07)

   gs |              1-loop Amplitude A1              
  ----------------------------------------------------
    0 | ( 0.00000000000000E+00, 0.00000000000000E+00)
    1 | ( 0.00000000000000E+00, 0.00000000000000E+00)
    2 | ( 0.14556032031458E-06, 0.82828108853855E-08)
    3 | ( 0.00000000000000E+00, 0.00000000000000E+00)
    4 | (-0.15273533002682E-05, 0.60662554073983E-05)
    5 | ( 0.00000000000000E+00, 0.00000000000000E+00)
  ----------------------------------------------------
  SUM | (-0.13817929799536E-05, 0.60745382182836E-05)   .
\end{verbatim}
}

The tables display the decomposition of the amplitude 
into contributions with different powers of the strong coupling
$g_s$ (as indicated in the first column).
The last line of the tables contains the sum of all the 
contributions, i.e.\ the full amplitude.

\item \texttt{writeMat = 2}:\\
  In addition to the output of \texttt{writeMat = 1}, \recola\ prints
  a decomposition of the one-loop amplitude (if computed) into the
  4-dimensional bare contribution (\texttt{4-dimensional bare-loop
    Amplitude A1d4}), the counterterm contribution (\texttt{CT
    Amplitude A1ct}) and the contribution of the rational parts $R_2$
  (\texttt{R2 Amplitude A1r2}), each of them again tabulated in powers
  of $g_s$:

{\footnotesize
\begin{verbatim}
   gs |    4-dimensional bare-loop Amplitude A1d4     
  ----------------------------------------------------
    0 | ( 0.00000000000000E+00, 0.00000000000000E+00)
    1 | ( 0.00000000000000E+00, 0.00000000000000E+00)
    2 | ( 0.13137505127039E-06, 0.84965453708879E-08)
    3 | ( 0.00000000000000E+00, 0.00000000000000E+00)
    4 | (-0.73297625052077E-06, 0.61361188439947E-05)
    5 | ( 0.00000000000000E+00, 0.00000000000000E+00)
  ----------------------------------------------------
  SUM | (-0.60160119925038E-06, 0.61446153893656E-05)

   gs |               CT Amplitude A1ct               
  ----------------------------------------------------
    0 | ( 0.00000000000000E+00, 0.00000000000000E+00)
    1 | ( 0.00000000000000E+00, 0.00000000000000E+00)
    2 | ( 0.11671303642281E-09,-0.14787829304366E-08)
    3 | ( 0.00000000000000E+00, 0.00000000000000E+00)
    4 | ( 0.52939559203394E-22,-0.99261673506363E-23)
    5 | ( 0.00000000000000E+00, 0.00000000000000E+00)
  ----------------------------------------------------
  SUM | ( 0.11671303642286E-09,-0.14787829304366E-08)

   gs |               R2 Amplitude A1r2               
  ----------------------------------------------------
    0 | ( 0.00000000000000E+00, 0.00000000000000E+00)
    1 | ( 0.00000000000000E+00, 0.00000000000000E+00)
    2 | ( 0.14068556007768E-07, 0.12650484449342E-08)
    3 | ( 0.00000000000000E+00, 0.00000000000000E+00)
    4 | (-0.79437704974740E-06,-0.69863436596445E-07)
    5 | ( 0.00000000000000E+00, 0.00000000000000E+00)
  ----------------------------------------------------
  SUM | (-0.78030849373963E-06,-0.68598388151511E-07)   .
\end{verbatim}
}%
\eei

\subsubsection*{Output options for the squared amplitude}

The value of the variable \texttt{writeMat2} determines the level of detail 
at which the results for 
the squared amplitudes $\big(\,\overline{\!{\cal A}^2\!}\,\big)_{\!0}$ 
and $\big(\,\overline{\!{\cal A}^2\!}\,\big)_{\!1}$ (defined in 
\refse{squared amplitudes}) are printed for all computed processes:

\bei

\item \texttt{writeMat2 = 0}:\\
Nothing is printed.

\item \texttt{writeMat2 = 1}:\\
The result for $\big(\,\overline{\!{\cal A}^2\!}\,\big)_{\!0}$,
denoted as $\texttt{| A0 |\^{}2}$,
and the result for $\big(\,\overline{\!{\cal A}^2\!}\,\big)_{\!1}$ (if computed),
denoted as $\texttt{2*Re\{ A1 * A0\^{}* \}}$ in the case of an existing
tree-level contribution and as 
$\texttt{| A1 |\^{}2}$ in the case of a loop-induced process,
are written in separate tables:

{\footnotesize
\begin{verbatim}
UNPOLARIZED SQUARED AMPLITUDE [GeV^-2]

 als |        | A0 |^2                als |   2*Re{ A1 * A0^* }   
-----------------------------        -----------------------------
   0 |  0.00000000000000E+00            0 |  0.00000000000000E+00
   1 |  0.00000000000000E+00            1 |  0.00000000000000E+00
   2 |  0.28817412651700E-06            2 | -0.12242757592358E-06
   3 |  0.00000000000000E+00            3 | -0.89053802197584E-06
     |                                  4 |  0.00000000000000E+00
-----------------------------        -----------------------------
 SUM |  0.28817412651700E-06          SUM | -0.10129655978994E-05  .
\end{verbatim}
}

The tables display the decomposition of the amplitude 
into contributions with different powers of the strong coupling
$\alpha_s$ (as indicated in the first column).
The last line of the tables contains the sum of all the 
contributions, i.e.\ the full squared amplitude.

\item \texttt{writeMat2 = 2}:\\
The same output as in the case \texttt{writeMat2 = 1} is returned, extended by 
a decomposition of the squared one-loop amplitude $\big(\,\overline{\!{\cal A}^2\!}\,\big)_{\!1}$
(if computed) into
the 4-dimensional bare-loop contribution 
(\texttt{2*Re\{A1d4*A0\^{}*\}}), the counterterm 
contribution (\texttt{2*Re\{A1ct*A0\^{}*\}}) and the contribution of 
the rational parts $R_2$ (\texttt{2*Re\{A1r2*A0\^{}*\}}), 
each of them again tabulated in powers of $\alpha_s$:

{\footnotesize
\begin{verbatim}
 als |   2*Re{A1d4*A0^*}  |   2*Re{A1ct*A0^*}  |   2*Re{A1r2*A0^*}  
-------------------------------------------------------------------
   0 |  0.00000000000E+00 |  0.00000000000E+00 |  0.00000000000E+00
   1 |  0.00000000000E+00 |  0.00000000000E+00 |  0.00000000000E+00
   2 | -0.11688477171E-06 | -0.14775782556E-08 | -0.40652259615E-08
   3 | -0.86084967845E-06 | -0.10229117946E-24 | -0.29688343530E-07
   4 |  0.00000000000E+00 |  0.00000000000E+00 |  0.00000000000E+00
-------------------------------------------------------------------
 SUM | -0.97773445015E-06 | -0.14775782556E-08 | -0.33753569492E-07 .
\end{verbatim}
}

For loop-induced processes, six columns are printed containing the
three squared contributions \texttt{|A1d4|\^{}2}, \texttt{|A1ct|\^{}2}
and \texttt{|A1r2|\^{}2}, and the three interference terms
\texttt{2*Re\{A1d4*A1ct\^{}*\}}, \texttt{2*Re\{A1ct*A1r2\^{}*\}} and
\texttt{2*Re\{A1r2*A1d4\^{}*\}}.

\item \texttt{writeMat2 = 3}:\\
  The same output as in the case \texttt{writeMat2 = 1} is returned,
  extended by a decomposition into the contributions $\overline{{\cal
      A}_h^2}$ from the single helicity configurations $\vec{h}$.  In
  this case, the (averaged) sum is only performed over colours.  Only
  helicity configurations with non-vanishing contributions are
  displayed (for the notation see the description for \texttt{writeMat
    = 1}):

{\footnotesize
\begin{verbatim}
POLARIZED SQUARED AMPLITUDE [GeV^-2]

Helicity configuration:   u[+] u~[-] -> Z[+] g[+] g[+]

 als |       | A0h |^2                als |  2*Re{ A1h * A0h^* }  
-----------------------------        -----------------------------
   0 |  0.00000000000000E+00            0 |  0.00000000000000E+00
   1 |  0.00000000000000E+00            1 |  0.00000000000000E+00
   2 |  0.28191064190249E-15            2 | -0.81359616022317E-15
   3 |  0.00000000000000E+00            3 | -0.98362830083235E-13
     |                                  4 |  0.00000000000000E+00
-----------------------------        -----------------------------
 SUM |  0.28191064190249E-15         SUM  | -0.99176426243458E-13  .
\end{verbatim}
}
\eei

\subsection*{Output options for spin- or colour-correlated squared amplitudes
}

The value of the variable \texttt{writeCor} determines the level of detail 
at which the results for the colour- and/or spin-correlated squared
tree amplitudes (see \refse{correlation}) are printed for all computed
processes:
\bei

\item \texttt{writeCor = 0}:\\
Nothing is printed.

\item \texttt{writeCor = 1}:\\
The colour-correlated squared Born amplitude 
$\big(\,\overline{\!{\cal A}^2\!}\,\big)_{\!\mathrm{c}}(n,m)$ between all 
pairs of particles $n$ and $m$ is shown, denoted as
\texttt{| A0c(}$n$\texttt{,}$m$\texttt{) |\^{}2}. 
For instance, for the colour correlation between particle $1$ and $3$ in the 
process $\mathrm{g g} \to \mathrm{d \bar{d} e^+ e^-}$ the sample output reads:

{\footnotesize
\begin{verbatim}
  COLOUR-CORRELATED SQUARED AMPLITUDE [GeV^-4]

  als |    | A0c(1,3) |^2    
  ----------------------------
    0 |  0.00000000000000E+00
    1 |  0.00000000000000E+00
    2 | -0.73462206785491E-11
    3 |  0.00000000000000E+00
    4 |  0.00000000000000E+00
  ----------------------------
  SUM | -0.73462206785491E-11   .
\end{verbatim}
}

The spin- and colour-correlated squared Born amplitude
$\big(\,\overline{\!{\cal A}^2\!}\,\big)_{\!\mathrm{sc}}(n,m,v)$ 
between a gluon $n$ with special polarization vector $v$ and an 
arbitrary coloured particle $m$ is shown, denoted as
\texttt{| A0sc(}$n$\texttt{,}$m$\texttt{) |\^{}2}. 
For the example from above, it reads: 

{\footnotesize
\begin{verbatim}
  SPIN- AND COLOUR-CORRELATED SQUARED AMPLITUDE [GeV^-4]

  Polarization vector v for particle 1:
  v(0) =  ( -1151.5040504618346     ,  0.0000000000000000     )
  v(1) =  ( -497.28644383654944     ,  0.0000000000000000     )
  v(2) =  (  580.15008908958009     ,  0.0000000000000000     )
  v(3) =  ( -1151.5040504618348     ,  0.0000000000000000     )

  als |   | A0sc(1,3) |^2    
  ----------------------------
    0 |  0.00000000000000E+00
    1 |  0.00000000000000E+00
    2 | -0.16797878275849E-05
    3 |  0.00000000000000E+00
    4 |  0.00000000000000E+00
  ----------------------------
  SUM | -0.16797878275849E-05   ,
\end{verbatim}
}

where \texttt{v(0:3)} is the polarization vector introduced by the 
user for the gluon $n$.

The spin-correlated squared Born amplitude 
$\big(\,\overline{\!{\cal A}^2\!}\,\big)_{\!\mathrm{s}}(n)$ for a photon $n$
is shown, denoted as \texttt{| A0s(}$n$\texttt{) |\^{}2}. 
For instance, for the spin correlation of the first photon in the process 
$\gamma \gamma \to \mu^+ \nu^- \mathrm{e^+ e^-}$ the sample output reads:

{\footnotesize
\begin{verbatim}
  SPIN-CORRELATED SQUARED AMPLITUDE [GeV^-4]

  Polarization vector v for particle 1:
  v(0) =  ( -1151.5040504618346     ,  0.0000000000000000     )
  v(1) =  ( -497.28644383654944     ,  0.0000000000000000     )
  v(2) =  (  580.15008908958009     ,  0.0000000000000000     )
  v(3) =  ( -1151.5040504618348     ,  0.0000000000000000     )

  als |     | A0s(1) |^2    
  ----------------------------
    0 |  0.84005617404439E-06
    1 |  0.00000000000000E+00
    2 |  0.00000000000000E+00
    3 |  0.00000000000000E+00
    4 |  0.00000000000000E+00
  ----------------------------
  SUM |  0.84005617404439E-06   .
\end{verbatim}
}
\eei
By default, the output is switched off completely:
\begin{verbatim}
  integer :: writeMat  = 0
  integer :: writeMat2 = 0
  integer :: writeCor  = 0 .
\end{verbatim}

\subsubsection*{Output options for memory needed by \recola.}

\recola\ can print information on the amount of memory (RAM)
needed in the present run. If and to which extent 
this information is written, is determined by the variable \texttt{writeRAM}:
\bei
\item \texttt{writeRAM = 0}:\\
Nothing is printed.
\item \texttt{writeRAM = 1}:\\
The amount of RAM permanently blocked by the 
information stored during the process generation, 
and the amount of RAM needed for the process computation 
are written to the standard output channel. In both cases
the information is written before the memory is actually occupied,
so that it can be used in the case of a memory overflow to detect the origin.
The amount of RAM for process computation is already shown during
the process generation, i.e.\ before any future call of 
\texttt{compute\_process\_rcl} that could exceed the memory limit of 
the computer. 

\item \texttt{writeRAM = 2}:\\
The same output as in the case \texttt{writeRAM=1} is returned, and in
addition also the amount of RAM temporally blocked during the process
generation is printed.  This memory is freed again before the end of
the generation step.
\eei

By default nothing is printed:
\begin{verbatim}
  integer :: writeRAM  = 0 .
\end{verbatim}


\subsection{Input subroutines}
\label{input subroutines}

The variables declared in the file \texttt{input.f90} and described in
detail in the previous section can be modified by the user also during
run time.  This is done with the help of the following input
subroutines (defined in \texttt{input.f90}), which can be called at
any time before the process generation is prompted via the call of
\texttt{generate\_processes\_rcl}.  
For most input subroutines, later calls have no effect unless \recola\
is reset (see \refse{reset}). Only certain subroutines can also be
employed at later stages, among them most notably the subroutine
{\tt{set\_alphas\_rcl}} which can be used to work with a dynamical
renormalization scale for the strong coupling $\alpha_s$. In the
following descriptions, the possibility of calling a subroutine also
after the process generation will be emphasized explicitly, i.e.\
unless stated otherwise the subroutine can only be employed before the
process generation.

\subsubsection{\tt{set\_pole\_mass\_{\it prtcl}\_rcl (m,g)}}

This subroutine sets the pole mass \texttt{mass\_{\it p}} and width
\texttt{width\_{\it p}} (in GeV) of the particle {\it prtcl} 
to \m\ and \g, respectively (\m\ and \g\ are of type 
{\tt real(dp)}).
Here, {\it prtcl} and {\it p} can take the following values:
\begin{center}\begin{tabular}[b]{l|cccccccccc}
particle & Z 
& W 
& Higgs & muon & 
tauon & charm & top  & bottom \\ 
 &  & & 
& & & quark &  quark & quark \\ 
\hline
{\it prtcl} & {\texttt z} & {\texttt w} & {\texttt h} &{\texttt muon} & 

{\texttt tau} & {\texttt charm} & {\texttt top} & {\texttt bottom} \\ 
\hline
{\it p} & {\texttt z} & {\texttt w} & {\texttt h} & {\texttt mu} &
{\texttt ta} & {\texttt c} & {\texttt t} & {\texttt b} \\ 
\end{tabular}\;.
\end{center}

\subsubsection{\tt{set\_pole\_mass\_{\it prtcl}\_rcl (m)}}

This subroutine sets the pole mass \texttt{mass\_{\it p}} (in GeV) of
the particle {\it prtcl} to \m\ (of type {\tt real(dp)}).  Here, {\it
  prtcl} and {\it p} can take the following values:
\begin{center}\begin{tabular}[b]{l|cccccccccc}
particle & electron & up & down  & strange \\ 
         &          &   quark &  quark & quark \\
\hline 
{\it prtcl} & {\texttt electron} & {\texttt up} & {\texttt down} & {\texttt
  strange} \\
\hline
{\it p} & {\texttt el} & {\texttt u} & {\texttt d} & {\texttt s} 
\end{tabular}\;.
\end{center}

\subsubsection{\tt{set\_onshell\_mass\_{\it v}\_rcl (m,g)}}

This subroutine sets the on-shell mass and width (in GeV) of the W or 
Z boson, ${\it v}={\texttt w}, {\texttt z}$, to \m\ and \g, 
respectively (\m\ and \g\ are of type {\tt real(dp)}).
Since \recola\ internally works with pole masses, the values of \m\ 
and \g\ are used to set the pole mass \texttt{mass\_{\it v}} and the
pole width \texttt{width\_{\it v}} according to
\bq
\texttt{mass\_{\it v}} = \frac{\m}{\sqrt{1+\g^2/\m^2}},
\qquad\qquad
\texttt{width\_{\it v}} = \frac{\g}{\sqrt{1+\g^2/\m^2}}.
\eq

\subsubsection{\tt{set\_light\_fermions\_rcl (m)}}

This subroutine declares all massive fermions with masses (in GeV)
smal\-ler than or equal to \m\ (of type {\tt real(dp)}) as light
(i.e.\ the variable \texttt{light\_}$\,f$ is set to \texttt{.true.}\ 
for each fermion $f$ with \texttt{mass\_}$\,f \leq$ \m).  Massive
fermions with masses larger than \m\ are declared heavy (i.e.\ the
variable \texttt{light\_}$\,f$ is set to \texttt{.false.}\ for each
fermion $f$ with \texttt{mass\_}$\,f >$ \m).  The subroutine call has
no effect on massless fermions.  Details on the variables
\texttt{light\_}$\,f$ are given in \refse{input collinear}.  Note that
the assignments made by \texttt{set\_light\_fermions\_rcl} are not
adjusted in case the fermion masses should be changed later on.

\subsubsection{\tt{set\_light\_{\it ferm}\_rcl}, 
             \texttt{unset\_light\_{\it ferm}\_rcl}}

The first subroutine marks 
the fermion {\it ferm} as light (i.e.\ it sets 
\texttt{light\_$\,f$ = .true.}), the second as heavy 
(i.e.\ it sets \texttt{light\_$\,f$ = .false.}). 
Here {\it ferm} and {\it f} can take the following values:
\begin{center}\begin{tabular}{l|cccc}
fermion & electron & muon & tauon\\
\hline 
{\it ferm} & {\texttt electron} & {\texttt muon}  & {\texttt tau}  
\\
\hline
{\it f} & {\texttt el} & {\texttt mu} & {\texttt ta} 
\end{tabular}
\\[2ex]
\begin{tabular}[b]{l|ccccccc}
fermion &  
up & down  & charm & strange & top & bottom 
\\ 
      & 
quark &  quark & quark & quark &  quark & quark \\
\hline 
{\it ferm} &
{\texttt up} & {\texttt down}  & {\texttt charm}
 &  {\texttt strange}  &  {\texttt top}  &  {\texttt bottom} 
\\
\hline
{\it f} & {\texttt u} & {\texttt d} & {\texttt c}  & {\texttt s} & {\texttt t} & {\texttt b} 
\end{tabular}\;.
\end{center}
If the fermion {\it ferm} is massless the subroutine call has no
effect.  Details on the variables \texttt{light\_}$\,f$ are given in
\refse{input collinear}.

\subsubsection{\tt{\tt use\_dim\_reg\_soft\_rcl}}

This subroutine selects dimensional regularization for soft
singularities (i.e.\ it sets \texttt{reg\_soft = 1}).  See
\refse{input soft} for details.

\subsubsection{\tt{use\_mass\_reg\_soft\_rcl (m)}}

This subroutine selects mass regularization for soft singularities
(i.e.\ it sets \texttt{reg\_soft = 2}), and it sets (in GeV) the mass
regulator to \m\ (of type {\tt real(dp)}).  See \refse{input soft} for
details.

\subsubsection{\tt{set\_delta\_uv\_rcl (d)}}

This subroutine sets the variable \deltauv\ parametrizing the UV pole
to \dA\ (of type {\tt real(dp)}).  See \refse{input dimensional
  regularization} for details.

\subsubsection{\tt{set\_mu\_uv\_rcl (m)}}

This subroutine sets (in GeV) the UV scale \muuv\ to \m\ 
(of type {\tt real(dp)}).  See \refse{input dimensional
  regularization} for details.

\subsubsection{\tt{set\_delta\_ir\_rcl (d,d2)}}

This subroutine sets the variables \deltair\ and \deltairr\ 
parametrizing the IR poles to \dA\ and to \dB, respectively (\dA\ and
\dB\ are of type {\tt real(dp)}).  See \refse{input dimensional
  regularization} for details.

\subsubsection{\tt{set\_mu\_ir\_rcl (m)}}

This subroutine sets (in GeV) the IR scale \muir\ to \m\ (of type {\tt
  real(dp)}).  See \refse{input dimensional regularization} for
details.

\subsubsection{\tt{set\_alphas\_rcl (a,s,Nf)}}

This subroutine sets the value of $\alpha_s$ (\als) to \aaa, the
renormalization scale \Qren\ to \s\ and the number of light quark
flavours \Nfren\ to \Nf\ (\aaa\ and \s\ are of type {\tt real(dp)},
\Nf\ is of of type {\tt integer}).  Details on the parameters \als,
\Qren\ and \Nfren\ can be found \refse{input alphas}.  
The supplied value of \aaa\ should correspond to the physical value of
$\alpha_s$ at the scale \s\ in the \Nf-flavour scheme.  Alternatively,
\recola\ can take care of the determination of the corresponding
$\alpha_s$ by means of the subroutine
\texttt{compute\_running\_alphas\_rcl} introduced in \refse{corual}.
The subroutine \texttt{set\_alphas\_rcl (a,s,Nf)} can be called before
the process generation but also during the process computation, which
allows to work with running values for $\alpha_s$.

\subsubsection{\tt{get\_alphas\_rcl (a)}}

This subroutine, which can be called before the process generation but
also during the process computation, returns the current value of
$\alpha_s$ in its output argument \texttt{a} (of type {\tt real(dp)}).

\subsubsection{\tt{get\_renormalization\_scale\_rcl (mu)}}

This subroutine, which can be called before the process generation but
also during the process computation, returns the current value of the
renormalization scale $Q$ (in GeV) for $\alpha_s$ in its output
argument \texttt{mu} (of type {\tt real(dp)}).

\subsubsection{\tt{get\_flavour\_scheme\_rcl (Nf)}}

This subroutine, which can be called before the process generation but
also during the process computation, returns the current value of the
identifier of the flavour-scheme $\Nfren$ for the renormalization of
$\alpha_s$ in its output argument \texttt{Nf} (of type {\tt integer}).
See \refse{input alphas} for details. 

\subsubsection{\tt{use\_gfermi\_scheme\_rcl (g)}}

This subroutine selects the $G_{\rm F}$ scheme as renormalization
scheme for the EW coupling. If the optional argument \g\ is present,
the Fermi constant $G_{\rm F}$ (\gf) is set to the value of \g\ (of
type {\tt real(dp)}).  See \refses{ew renormalization} and \ref{input
  ew renormalization} for
details.

\subsubsection{\tt{use\_alpha0\_scheme\_rcl (a)}}

This subroutine selects the $\alpha(0)$ scheme as renormalization
scheme for the EW coupling. If the optional argument \aaa\ is present,
$\alpha(0)$ (\alO) is set to the value of \aaa\ (of type {\tt
  real(dp)}).  See \refses{ew renormalization} and \ref{input ew
  renormalization} for details.

\subsubsection{\tt{use\_alphaz\_scheme\_rcl (a)}}

This subroutine selects the $\alpha(M_{\rm Z})$ scheme as
renormalization scheme for the EW coupling. If the optional argument
\aaa\ is present, $\alpha(M_{\rm Z})$ (\alZ) is set to the value of
\aaa\ (of type {\tt real(dp)}).  See \refses{ew renormalization} and
\ref{input ew renormalization} for details.

\subsubsection{\tt{get\_alpha\_rcl (a)}}

This subroutine, which can be called before the process generation but
also during the process computation, returns the current value of
$\alpha$ in its output argument \texttt{a} (of type {\tt real(dp)}).

\subsubsection{\tt{set\_complex\_mass\_scheme\_rcl}}

\begin{sloppypar}
  This subroutine selects the complex-mass scheme
  of~\citeres{Denner:1999gp,Denner:2005fg,Denner:2006ic} for the
  renormalization of the masses of unstable particles (i.e.\ it sets
  \texttt{complex\_mass\_scheme = 1}).  See \refse{input cms} for
  details.
\end{sloppypar}

\subsubsection{\tt{set\_on\_shell\_scheme\_rcl}}

\begin{sloppypar}
  This subroutine selects the on-shell scheme for the mass
  renormalization of unstable particles (i.e.\ it sets
  \texttt{complex\_mass\_scheme = 0}).  See \refse{input cms} for
  details.
\end{sloppypar} 

\subsubsection{\tt{set\_resonant\_particle\_rcl (pa)}}
\label{setrespart}

This subroutine is only needed if processes are defined 
with specific intermediate particles and the pole approximation shall be
applied to the corresponding resonances (see \refse{sec:defpro} and
\refse{resonances}). 
It marks particle \texttt{pa} as resonant. 
The argument \texttt{pa} can be any of the characters listed in 
\refse{conventions} for the SM particles. 

If a particle is labelled resonant, \recola\ sets the imaginary part
of its mass to zero everywhere except in the denominator of the
resonant propagators. If the width of the particle \texttt{pa} is
zero, \recola\ stops.  During computation, \recola\ furthermore checks
that the phase-space point provided by the user fulfils the condition
that the squared momenta of all resonant particles \texttt{pa} are on
shell (if this does not hold within a precision of $10^{-7}$, \recola\ 
stops).

Calling \texttt{set\_resonant\_particle\_rcl (pa)} affects all
\texttt{compute\_$X$\_rcl} calls of \refse{computation}, i.e.\
the pole approximation is also applied to colour- and/or
spin-correlated squared amplitudes.

\subsubsection{\tt{switchon\_resonant\_selfenergies\_rcl}}

This subroutine sets the value of \texttt{resSE} to \texttt{.true.}
(see \refse{input resSE} for details).

\subsubsection{\tt{switchoff\_resonant\_selfenergies\_rcl}}

This subroutine sets the value of \texttt{resSE} to \texttt{.false.}
(see \refse{input resSE} for details).

\subsubsection{\tt{set\_draw\_level\_branches\_rcl (n)}}

This subroutine sets the variable \texttt{draw} for the generation 
of a \LaTeX\ file with diagrams of the off-shell currents 
to \n\ (of type {\tt integer}).
See \refse{input draw} for details. 

\subsubsection{\tt{set\_print\_level\_amplitude\_rcl (n)}}

This subroutine, which can be called before the process generation but
also during the process computation, sets the variable
\texttt{writeMat} governing the output of amplitudes to \n\ (of type
{\tt integer}).  See \refse{input print} for details.

\subsubsection{\tt{set\_print\_level\_squared\_amplitude\_rcl (n)}}

This subroutine, which can be called before the process generation but
also during the process computation, sets the variable
\texttt{writeMat2} governing the output of squared amplitudes to \n\
(of type {\tt integer}).  See \refse{input print} for details.

\subsubsection{\tt{set\_print\_level\_correlations\_rcl (n)}}

\begin{sloppypar}
  This subroutine, which can be called before the process generation
  but also during the process computation, sets the variable
  \texttt{writeCor} governing the output of colour- and
  spin-correlated squared amplitudes to \n\ (of type {\tt integer}).
  See \refse{input print} for details.
\end{sloppypar}

\subsubsection{\tt{set\_print\_level\_RAM\_rcl (n)}}

This subroutine sets the variable \texttt{writeRAM} governing the
output of information on the amount of RAM needed to \n\ (of type {\tt
  integer}).  See \refse{input print} for details.

\subsubsection{\tt{scale\_coupling3\_rcl (fac,pa1,pa2,pa3)}}

This subroutine scales the 3-particle coupling between particles \paA,
\paB, and \paC\ by a factor \fac\ (of type {\tt complex(dp)}).  The
arguments \paA, \paB, and \paC\ can be any of the characters listed in
\refse{conventions} for the SM particles.

\begin{remark} 
The rescaling of couplings in the counterterms and rational terms is 
not yet implemented.
\end{remark}

\subsubsection{\tt{scale\_coupling4\_rcl (fac,pa1,pa2,pa3,pa4)}}

This subroutine scales the 4-particle coupling between particles \paA,
\paB, \paC, and \paD\ by a factor \fac\ (of type {\tt complex(dp)}).
The arguments \paA, \paB, \paC, and \paD\ can be any of the characters
listed in \refse{conventions} for the SM particles.

\begin{remark} 
The rescaling of couplings in the counterterms and rational terms is 
not yet implemented.
\end{remark}

\subsubsection{\tt{switchoff\_coupling3\_rcl (pa1,pa2,pa3)}}

This subroutine switches off the 3-particle coupling between \paA,
\paB, and \paC.  The arguments \paA, \paB, and \paC\ can be any of the
characters listed in \refse{conventions} for the SM particles.

\begin{remark}
  The switching off of couplings in the counterterms and rational
  terms is not yet implemented.
\end{remark}

\subsubsection{\tt{switchoff\_coupling4\_rcl (pa1,pa2,pa3,pa4)}}

This subroutine switches off the 4-particle coupling between \paA,
\paB, \paC, and \paD.  The arguments \paA, \paB, \paC, and \paD\ can
be any of the characters listed in \refse{conventions} for the SM
particles.

\begin{remark} 
The switching off of couplings in the counterterms and rational terms 
is not yet implemented.
\end{remark}

\subsubsection{\tt{set\_ifail\_rcl (i)}}

This subroutine, which can be called before the process generation but
also during the process computation, sets the flag \texttt{ifail} to
the {\tt integer} value \texttt{i} given by the user.  On input the
flag \texttt{ifail} determines the behaviour of \recola\ upon
encountering an internal error:
\bei
\item
\texttt{ifail = -1}: Execution of the program does not stop if an error occurs;
\item
\texttt{ifail = \;\,0}: Execution of the program stops if an error occurs.
\eei
By default, the flag \texttt{ifail} is set to \texttt{0}.

\subsubsection{\tt{get\_ifail\_rcl (i)}}

This subroutine, which can be called before the process generation but
also during the process computation, extracts the value of the flag
\texttt{ifail}, returning it as {\tt integer} argument {\tt i} to the
user.  On output, the flag \texttt{ifail} provides information about
the appearance of an error during the run of \recola.
\bei
\item
\texttt{ifail = 0}: No errors occurred;

\item
\texttt{ifail = 1}: An error occurred due to an inconsistent call of
a subroutine by the user;

\item
\texttt{ifail = 2}: An error occurred due to an unexpected problem within
\recola.
\eei

\subsubsection{\tt{set\_output\_file\_rcl (x)}}

This subroutine, which can be called before the process generation but
also during the process computation, opens the output file named
\texttt{x} (\texttt{x} is of type {\tt character}).  Any output
produced by \recola\ is written to the output file \texttt{x}, which
is created with the first call of \texttt{set\_output\_file\_rcl} (if
it already exists, its content is cleared). For any subsequent call of
\texttt{set\_output\_file\_rcl} with different argument a new
output file is opened, and if it exists its content is cleared.
Redirecting the output to an already open output file does not clear
the content and new output is appended.  Output files remain open
unless recola is reset by the call of \texttt{reset\_recola\_rcl}.

If the argument \texttt{x} = \texttt{'*'} is passed to
\texttt{set\_output\_file\_rcl}, the output is written into the
terminal (standard output channel).

By default, the output is written to the file \texttt{output.rcl},
placed in the current working directory. The call of
\texttt{reset\_recola\_rcl} resets the output file name to the default
value \texttt{output.rcl}.


\subsection{Process definition}
\label{definition}

Processes are declared by calling the subroutine
\texttt{define\_process\_rcl} defined in the file
\texttt{process\_definition.f90}. This file also contains subroutines
for the selection/unselection of contributions to the amplitude with
specific powers of the strong coupling $g_s$ (defined by $g_s^2 =
4\pi\alpha_s$). For a process with identifier {\tt npr}, these
subroutines can be called at any point after the declaration of the
process {\tt npr} via \texttt{define\_process\_rcl}, and before the
complete set of processes is generated by the call of
\texttt{generate\_processes\_rcl}.  If no specific selection is made
by the user, by default the contributions of all powers of $g_s$ are
computed.  Since the contributions to the amplitude are proportional
to $g_s^{n_s}e^{n-n_s}$ with $n$ unambiguously fixed from the chosen
process (and the selection of LO or NLO), the powers of the
electromagnetic coupling $e$ (defined by $e^2 = 4\pi\alpha$) are
implicitly determined by the powers of $g_s$.

\subsubsection{\tt{define\_process\_rcl (npr,processIn,order)}}
\label{sec:defpro}
With this subroutine, the user defines the process(es) he wants to be
computed.  The {\tt integer} argument \npr\ is assigned as identifier
to the process specified by the argument \processIn\ (of type {\tt
  character}).  Issuing repeated calls of
\texttt{define\_process\_rcl}, it is possible to define more than one
process. In this case, each process \processIn\ must be given a
different identifier \npr\ (calls of {\tt define\_process\_rcl} with
an already existing identifier are ignored).  The argument \order\ of
type {\tt character} can take the values \texttt{'LO'} and
\texttt{'NLO'}: For \texttt{\order='LO'}, the process will be
generated such that \recola\ will be able to evaluate only LO
amplitudes, while for \texttt{\order='NLO'} it will be generated such
that \recola\ will be able to evaluate both LO and NLO amplitudes.

The string \processIn\ must consist of a list of particles separated
by \texttt{'->'}, with the incoming (outgoing) particles on the
left-hand (right-hand) side of \texttt{'->'}.  Any number of incoming
and outgoing particles is allowed.  The symbols for the particles,
listed in \refse{conventions}, as well as the symbol \texttt{'->'}
must be separated by at least one blank character.  Examples of
allowed character chains for \processIn\ are
\begin{verbatim}
   'e+ e- -> mu+ mu-' ,
   'u u~ -> W+ W- g' ,
   'u d~ -> W+ g g g' ,
   'u  g  ->  u  g  Z' ,
   'u    g  -> u        g  tau- tau+' .
\end{verbatim}
To each fermion or vector particle 
a specific helicity can be attributed by adding the corresponding symbol 
\texttt{'[-]'}, \texttt{'[+]'} or \texttt{'[0]'} (see \refse{conventions} for 
the definition of the helicity symbols). 
Blank characters can (but need not) separate the helicity symbol 
from the particle name:
\begin{insertion}%
\begin{tabular}{@{}ll@{}}
\texttt{'e+[+] e- [-] -> Z H'}: &
\texttt{'e+'} is right-handed, \texttt{'e-'} is \rlap{left-handed,}
\\ &
\texttt{'Z'} is unpolarized.
\\
\texttt{'u u\~\ -> W+[-] W-[+]'}: & 
\texttt{'u'} and \texttt{'u\~\ $\!\!\!$'} are unpolarized, 
\\ &
\texttt{'W+'} and \texttt{'W-'} are transverse.
\end{tabular}%
\end{insertion}
Contributions with specific intermediate states can be selected (see
\refse{resonances}), where intermediate states are particles decaying
into any number of other particles.  To this end, in the process
declaration the decaying particle must be followed by round brackets
\mbox{\texttt{'( ... )'}} containing the decay products.  Multiple and
nested decays are allowed.  Blank characters can (but need not)
separate the brackets \texttt{'('}, \texttt{')'} from the particle
names:
\begin{verbatim}
   'e+ e- -> W+ W-(e- nu_e~)' ,
   'e+ e- -> Z H ( b~[+] b[-] )' ,
   'e+ e- -> t( W+(u d~) b) t~(e- nu_e~ b~)' ,
   'u  u~ -> Z ( mu+(e+ nu_e nu_mu~) mu-(e- nu_e~ nu_mu) ) H' .
\end{verbatim}
The selection of specific intermediate particles is a prerequisite for
the calculation of amplitudes within the pole approximation.

\subsubsection{\tt{set\_gs\_power\_rcl (npr,gsarray)}}

This subroutine selects specific powers of $g_s$, as 
specified by the argument \texttt{gsarray}, for the process with
identifier \npr\ (of type {\tt integer}).  
The elements of the
two-dimensional {\tt integer} array \texttt{gsarray}(0\,:\,,0\,:\,1) can take
the values 0 or 1, with the first index representing the power of
$g_s$ and the second one representing the loop order (0 for LO, 1 for
NLO).
For instance, a call of \texttt{set\_gs\_power\_rcl} with second argument
\begin{verbatim}
   gsarray(0,0) = 1
   gsarray(1,0) = 0
   gsarray(2,0) = 1
   gsarray(0,1) = 0
   gsarray(1,1) = 0
   gsarray(2,1) = 0
   gsarray(3,1) = 0
   gsarray(4,1) = 1 
\end{verbatim}
selects the contributions with $g_s^0$ and $g_s^2$ for the tree
amplitude and the $g_s^4$ contribution for the loop amplitude. 
All other contributions will not be computed in later evaluations 
of the amplitudes. 
By default, all contributions are switched on.

\subsubsection{\tt{select\_gs\_power\_BornAmpl\_rcl (npr,gspower)}, \\
               \texttt{unselect\_gs\_power\_BornAmpl\_rcl (npr,gspower)}}%

This pair of subroutines allows to select/unselect the contribution to 
the Born amplitude proportional to $g_s^n$, 
where $n$ is given by the {\tt integer} argument \gspower, 
for the process with identifier \npr\ (of type {\tt integer}).
All other contributions to the Born amplitude keep their status (selected or 
unselected), according to previous calls of selection subroutines. 
The selection of the contributions to the loop amplitude remains 
unaffected as well.

\subsubsection{\tt{select\_gs\_power\_LoopAmpl\_rcl (npr,gspower)}, \\
             \texttt{unselect\_gs\_power\_LoopAmpl\_rcl (npr,gspower)}}

This pair of subroutines allows 
to select/unselect the contribution to 
the loop amplitude proportional to $g_s^n$, where $n$ is given 
by the {\tt integer} argument \gspower, 
for the process with identifier \npr\ (of type {\tt integer}). 
All other contributions to the loop amplitude keep their status (selected or 
unselected), according to previous calls of selection 
subroutines. 
The selection of the contributions to the Born amplitude remains 
unaffected as well.

\subsubsection{\tt{select\_all\_gs\_powers\_BornAmpl\_rcl (npr)}, \\
             \texttt{unselect\_all\_gs\_powers\_BornAmpl\_rcl (npr)}}

This pair of subroutines allows to select/unselect all contributions to 
the Born amplitude (with any power of $g_s$) for the process with 
identifier \npr\ (of type {\tt integer}).
The selection of the contributions to the loop amplitude remains 
unaffected. 

\subsubsection{\tt{select\_all\_gs\_powers\_LoopAmpl\_rcl (npr)}, \\
             \texttt{unselect\_all\_gs\_powers\_LoopAmpl\_rcl (npr)}}

This pair of subroutines allows to select/unselect all contributions to 
the loop amplitude (with any power of $g_s$) for the process with 
identifier \npr\ (of type {\tt integer}).
The selection of the contributions to the Born amplitude remains 
unaffected.


\subsection{Process generation: \tt{generate\_processes\_rcl}}
\label{generation}

The file \texttt{process\_generation.f90} only contains the single
subroutine \texttt{generate\_processes\_rcl} for the generation of the
processes. This subroutine constructs the skeleton of the recursive
procedure for all processes previously defined by the user, and stores
it in global variables that later on are used by \recola\ for the
computation of amplitudes.  It must be called {\bf once} after all
processes are defined and before the subroutines for process
computation contained in \texttt{process\_computation.f90} can be
called.  It is typically called before the phase-space points are
generated.

\subsection{Process computation}
\label{computation}

{\sloppypar
The amplitude and the squared amplitude for a process are computed
by calling the subroutine \texttt{compute\_process\_rcl}. 
Their values are stored in internal variables labelled by the process 
identifier and the loop order and can be read out with the subroutines 
\texttt{get\_amplitude\_rcl}, 
\texttt{get\_polarized\_squared\_amplitude\_rcl}
and \texttt{get\_squared\_amplitude\_rcl}. 
After the computation, the amplitudes and squared amplitudes can be 
rescaled for a new value of $\alpha_s$ with the subroutine 
\texttt{rescale\_process\_rcl} (which overwrites the stored results 
for the given process at the given order). 
The typical sequence is
\begin{verbatim}
  call compute_process_rcl(npr,...)
  call get_A_rcl(npr,...) ,
\end{verbatim}
possibly followed (after a redefinition of $\alpha_s$ through 
\texttt{set\_alphas\_rcl} or \texttt{compute\_running\_alphas\_rcl}) 
by several calls of
\begin{verbatim}
  call rescale_process_rcl(npr,...)
  call get_A_rcl(npr,...) ,
\end{verbatim}
where \texttt{npr} is the process identifier and \texttt{$A$} stands 
for \texttt{amplitude}, \texttt{polarized\_squared\_amplitude} and/or 
\texttt{squared\_amplitude}.

Further subroutines of type \texttt{compute\_$X$\_correlation\_rcl}
and \texttt{rescale\_$X$\_correlation\_rcl} (where \texttt{$X$} stands
for \texttt{colour}, \texttt{spin} or \texttt{spin\_colour}) are
provided for the computation of spin- and/or colour-correlated squared
Born amplitudes.  For colour correlation also the routines
\texttt{compute\_all\_colour\_correlations\_rcl} and
\texttt{rescale\_all\_colour\_correlations\_rcl} are available.  These
subroutines compute (or rescale) the LO amplitudes and the LO spin-
and/or colour-correlated squared amplitudes, storing the results in
internal variables (for LO amplitudes these variables are the same as
for \texttt{compute\_process\_rcl} and
\texttt{rescale\_process\_rcl}).  Previously computed results of these
LO objects for the given process are overwritten.  The stored results
for the spin- and/or colour-correlated squared Born amplitudes can be
read out by the user with the subroutines of type
\texttt{get\_$X$\_correlation\_rcl}.  The typical sequence for the
evaluation of spin- and/or colour-correlated squared Born amplitudes
is
\begin{verbatim}
  call compute_X_correlation_rcl(npr,...)
  call get_X_correlation_rcl(npr,...) ,
\end{verbatim}
possibly followed (after a redefinition of $\alpha_s$) by several calls
\begin{verbatim}
  call rescale_X_correlation_rcl(npr,...)
  call get_X_correlation_rcl(npr,...)
\end{verbatim}
for rescaling the LO amplitudes and spin- and/or colour-correlated amplitudes.

Note that a call of \texttt{compute\_$Y$\_rcl} or
\texttt{rescale\_$Y$\_rcl} followed by a call of
\texttt{compute\_$Y'$\_rcl} or \texttt{rescale\_$Y'$\_rcl} (with
\texttt{$Y$},\texttt{$Y'$} $=$ \texttt{process},
\texttt{$X$\_correlation}, \texttt{all\_colour\_correlations}) will
lead to a recalculation of LO amplitudes for process \texttt{npr} that
overwrites results previously obtained in the calls for $Y$ for that
process.  In this case the results for the amplitudes calculated with
\texttt{compute\_$Y$\_rcl} or \texttt{rescale\_$Y$\_rcl} cannot be
accessed anymore.  It is thus advisable to conclude the call sequence
for an object $Y$, before a different object $Y'$ is evaluated.  More
details are given in the description of the individual subroutines.

All these subroutines are defined in the file 
\sloppy
\texttt{process\_computation.f90}, together with the subroutine 
\texttt{compute\_running\_alphas\_rcl} that allows for the dynamical 
computation of $\alpha_s$ and the subroutine 
\texttt{set\_resonant\_squared\_momentum\_rcl} which enables the user 
to set the squared momentum of the denominator of resonant 
propagators.
}

\subsubsection{\tt{set\_resonant\_squared\_momentum\_rcl (npr,res,ps)}}
\label{resonantmomentum}

This subroutine is only relevant for processes defined with
intermediate particles (see \refse{sec:defpro} and
\refse{resonances}).  It acts on the resonance with identifier \res\
(of type {\tt integer}) in the process with identifier \npr\ (of type
{\tt integer}) in the following way:
\bei
\item
it checks whether the resonant particle corresponding to \res\ has
been marked resonant through the subroutine
\texttt{set\_resonant\_particle\_rcl} (see \refse{setrespart});
\item
it sets the squared momentum in the denominator of the resonant 
propagator to \ps\ (of type {\tt real(dp)}).  

\eei
The resonance identifier \res\ is derived from the process definition 
in the call of \texttt{define\_process\_rcl}. 
The first resonant particle defined there is labelled with \res\ = 1, 
the second with \res\ = 2, and so on.  For example, in the process 
\begin{verbatim}
   'e+ e- -> t( W+(u d~) b) t~(e- nu_e~ b~)' ,
\end{verbatim}
the first resonant particle (\res\ = 1) is the top-quark, the second
(\res\ = 2) the $\PW^+$, and the third (\res\ = 3) the anti-top quark.
In order to take effect on the calculation, this subroutine must be
called before \texttt{compute\_process\_rcl}.

\subsubsection{\tt{compute\_running\_alphas\_rcl (Q,Nf,lp)}}
\label{corual}

This subroutine can be called to compute the value for $\alpha_s$ at
the scale \texttt{Q} employing the renormalization-group evolution at
\texttt{lp} loops (\texttt{Q} is of type {\tt real(dp)}, the {\tt
  integer} argument \texttt{lp} can take the value \texttt{lp$=$1} or
\texttt{lp$=$2}).  The {\tt integer} argument \texttt{Nf} selects the
flavour scheme according to the following rules (see also
\refses{dimensional regularization} and \ref{alphas}):
\bei
\item\texttt{Nf = -1}:\\
The variable flavour scheme is selected, where the number of active
flavours contributing to the running of $\alpha_s$ is set to the
number of quarks lighter than the scale \texttt{Q}. 

\item\texttt{Nf = 3,4,5,6}:\\
The fixed $N_{\mathrm{f}}$-flavour scheme is selected, where the
number of active  flavours contributing to the running of $\alpha_s$
is set to $N_{\mathrm{f}}=\texttt{Nf}$.  In this case \texttt{Nf}
cannot be smaller than the number of massless quarks (otherwise the
code stops).
\eei
By calling the subroutine \texttt{compute\_running\_alphas\_rcl} the
new value of $\alpha_s$ is internally computed by \recola\ according
to the formulas given in \refse{alphas}.  Using this subroutine (or
alternatively \texttt{set\_alphas\_rcl}) after the call of
\texttt{generate\_processes\_rcl}, the user can assign a different
value to $\alpha_s$ at each phase-space point.

\subsubsection{\tt{compute\_process\_rcl (npr,p,order,A2)}} 
\label{compute-process}

{\sloppy
This subroutine computes the Born contribution 
${\cal A}_0^{(\vec{c},\vec{h})}$ and the one-loop contribution 
${\cal A}_1^{(\vec{c},\vec{h})}$ to the structure-dressed amplitude 
${\cal A}^{(\vec{c},\vec{h})}$ (see \refse{amplitude structure}) of the 
process with identifier \npr\ (of type {\tt integer}) for all values
of $\vec{c}$ and $\vec{h}$. It further computes the squared amplitudes
$\big(\,\overline{\!{\cal A}^2\!}\,\big)_{\!0}$ and
$\big(\,\overline{\!{\cal A}^2\!}\,\big)_{\!1}$ as defined in
\refse{squared amplitudes}.  
The argument \texttt{order} (of type \texttt{character}) can take the 
two values \texttt{'LO'} and \texttt{'NLO'}. 
The one-loop contributions are only
computed if \texttt{order='NLO'} and if the process has been defined 
at NLO in the call of \texttt{define\_process\_rcl} 
(see \refse{sec:defpro}).  
In the computation of the squared amplitude a sum/average is performed
over helicities and colours; for particles that have been defined with
a specific helicity, no helicity sum/average is carried out.  The
results for the amplitudes and squared amplitudes are stored in
internal variables (overwriting previous computed results for
amplitudes and squared amplitudes for process \npr\ up to order
\texttt{order}) that can be read out by the user with the subroutines
\texttt{get\_amplitude\_rcl},
\texttt{get\_polarized\_squared\_amplitude\_rcl} and
\texttt{get\_squared\_amplitude\_rcl}.

The input array \texttt{p(0:3,1:}$l$\texttt{)} of type {\tt real(dp)}
should be filled with the momenta $p_i^\mu$ of the $l$ external
particles of the process \npr.  The first index of \p\ refers to the
Lorentz index $\mu$, the second to the particle index $i$ of
$p^\mu_i$, with the particles ordered according to their position in
the process definition. The optional output argument \texttt{A2(0:1)}
is a vector of type {\tt real(dp)}.  Its two entries, \texttt{A2(0)}
and \texttt{A2(1)}, return the values for $\big(\,\overline{\!{\cal
    A}^2\!}\,\big)_{\!0}$ and $\big(\,\overline{\!{\cal
    A}^2\!}\,\big)_{\!1}$, respectively, with the contributions from
all selected powers of $\alpha_s$ summed up.  }

\subsubsection{\tt{rescale\_process\_rcl (npr,order,A2)}}
\label{rescale-process}

{\sloppy
This subroutine allows to adjust the results calculated by
\texttt{compute\_process\_rcl} for a new value of $\alpha_s$, without
recomputing the amplitudes (this is done by rescaling LO and NLO
amplitudes and recomputing the counterterm for the strong coupling).
The user first calls \texttt{compute\_process\_rcl} for a process 
with identifier \npr\ (of type {\tt integer}) at order 
\texttt{order} (of type \texttt{character} taking the 
two values \texttt{'LO'} or \texttt{'NLO'}), then sets a new value 
for $\alpha_s$ (by means of 
\texttt{set\_alphas\_rcl} or \texttt{compute\_running\_alphas\_rcl})
and finally calls \texttt{rescale\_process\_rcl} with the same 
process identifier \npr.

The call of \texttt{rescale\_process\_rcl} leads to rescaling of the
\sloppy stored results for amplitudes and squared amplitudes for
process \npr\ up to order \texttt{order} and overwrites previous results
({\tt order='NLO'}
requires that the original call of
\texttt{compute\_process\_rcl} had been evaluated at NLO).

The adjusted results can be obtained with help of the subroutines
\texttt{get\_amplitude\_rcl},
\texttt{get\_polarized\_squared\_amplitude\_rcl} and
\texttt{get\_squared\_amplitude\_rcl} or via the optional output
argument \texttt{A2(0:1)} (defined as for
\texttt{compute\_process\_rcl}).
}

\subsubsection{\tt{get\_amplitude\_rcl (npr,pow,order,colour,hel,A)}}
\label{get amplitude}

This subroutine extracts a specific contribution to the
structure-dressed amplitude ${\cal A}_0^{(\vec{c},\vec{h})}$ or ${\cal
  A}_1^{(\vec{c},\vec{h})}$ (see \refse{amplitude structure}) of the
process with identifier \npr\ (of type {\tt integer}), according to
the values of the arguments \pow, \order, \texttt{colour} and \hel:
\bei
\item 
\pow\ is of type {\tt integer} and specifies the power of $g_s$ of the 
contribution. 
\item \order\ represents the loop order of the contribution.  It is a
  variable of type {\tt character} accepting precisely the following
  values:
\begin{insertion}
\begin{tabular}{@{}ll@{}}
\texttt{'LO'}:     & Born amplitude, \\
\texttt{'NLO'}:    & complete one-loop amplitude, \\
\texttt{'NLO-D4'}: & bare 4-dimensional one-loop amplitude, \\
\texttt{'NLO-CT'}: & counterterm contribution to the one-loop \rlap{amplitude,}\\
\texttt{'NLO-R2'}: & rational-term contribution to the one-loop \rlap{amplitude.}
\end{tabular}
\end{insertion}
\item
\texttt{colour(1:$l$)} describes the colour structure $\vec{c}$
and is a vector of type {\tt integer} 
and length $l$, where each position in the vector
corresponds to one of the $l$ external particles of process \npr\ 
(ordered as in the process definition).
For colourless particles, incoming quarks and outgoing anti-quarks,
the corresponding entry in \texttt{colour} must be 0.
For all other particles (gluons, outgoing quarks and incoming anti-quarks),
the entries must be, without repetition, the positions 
of the gluons, incoming quarks and outgoing anti-quarks in the process 
definition. 
If $k_1,k_2,\dots,k_n$ ($n\le l$) are the positions in \texttt{colour}
that contain non-zero entries, and $m_1,m_2,\dots,m_n$ are the respective entries,
then \texttt{colour} represents the colour structure  
$\delta^{i_{k_1}}_{j_{m_1}}\,\delta^{i_{k_2}}_{j_{m_2}}\,\cdots\,\delta^{i_{k_n}}_{j_{m_n}}$.

\begin{example}
Process:   \texttt{'u u\~{} -> Z g g'}\\
Position: \;\texttt{1 2 \ \ \ \ 3 4 5} \;.
\end{example}

The vector \texttt{colour} has length 5.  The 1$^{\rm st}$ and the
3$^{\rm rd}$ entry corresponding to the up quark and the Z boson must
be 0 (incoming quark and colourless particle).  The other entries
(2$^{\rm nd}$, 4$^{\rm th}$ and 5$^{\rm th}$) must be filled with a
permutation of the values $\{1,4,5\}$, i.e.\ the positions of the
incoming up quark and the gluons.

Hence, a possible \texttt{colour} vector is 
\begin{insertion}
\texttt{colour(1:5) = [0,1,0,5,4]},
\end{insertion}
corresponding to the colour structure 
$\delta^{i_2}_{j_1}\,\delta^{i_4}_{j_5}\,\delta^{i_5}_{j_4}$. 

\item \hel\texttt{(1:$l$)} describes the helicity configuration
  $\vec{h}$ and is a vector of type {\tt integer} and length $l$,
  where each position in the vector corresponds to one of the $l$
  external particles of process \npr\ (ordered as in the process
  definition).  Its entries represent the helicities of the
  corresponding particles, according to the conventions of
  \refse{conventions}.

\begin{example}
Process: \texttt{'u g -> W+ d'}
\end{example}

A possible \hel\ vector is 
\begin{insertion}
\texttt{hel(1:4) = [-1,+1,-1,-1]} 
\end{insertion}
corresponding to the following helicities:
\begin{insertion}
\begin{tabular}{@{}ll}
up~quark   & $\to\quad$ helicity = $-1$, \\
gluon      & $\to\quad$ helicity = $+1$, \\
$\PW^+$    & $\to\quad$ helicity = $-1$, \\
down~quark & $\to\quad$ helicity = $-1$.
\end{tabular}%
\end{insertion}%
\eei
The argument \texttt{A} of type {\tt complex(dp)} delivers the output
consisting of the value of the amplitude for process \npr\ as stored
in internal variables. Any call of a subroutine of type
\texttt{compute\_$X$\_rcl} or \texttt{rescale\_$X$\_rcl} with the same
process identifier \npr\ (partially) overwrites these internal
variables. In order to extract the results from the last process
computation, the subroutine \texttt{get\_amplitude\_rcl} should thus
be called after \texttt{compute\_process\_rcl} or
\texttt{rescale\_process\_rcl} and before a different subroutine call
of type \texttt{compute\_$Y$\_rcl} or \texttt{rescale\_$Y$\_rcl} is
issued for the same process \npr. When called at  {\tt order$=$'LO'}
after \texttt{compute\_$Y$\_rcl} or \texttt{rescale\_$Y$\_rcl} with
$Y$ being \texttt{colour\_correlation}, \texttt{spin\_correlation},
\texttt{spin\_colour\_correlation} or
\texttt{all\_colour\_correlations}, the subroutine
\texttt{get\_amplitude\_rcl}  will return the LO matrix element
entering the calculation of the corresponding correlated squared
amplitude.

\subsubsection{\tt{get\_squared\_amplitude\_rcl (npr,pow,order,A2)}}
\label{get squared}

This subroutine extracts the computed value of the squared amplitude
$\big(\,\overline{\!{\cal A}^2\!}\,\big)_{\!0}$ or
$\big(\,\overline{\!{\cal A}^2\!}\,\big)_{\!1}$ (see \refse{squared
  amplitudes} and \refse{compute-process} for details on the
definition of squared amplitudes) for the process with identifier
\npr\ (of type {\tt integer}), according to the values of the
arguments \pow\ and \order: \bei
\item 
\pow\ is of type {\tt integer} and specifies the power of $\alpha_s$ of the 
contribution. 

\item
\order\ represents the loop order of the contribution. 
It is variable of type {\tt character} accepting the following values:
\begin{insertion}
\begin{tabular}{@{}ll@{}}
\texttt{'LO'}:     & squared Born amplitude, \\
\texttt{'NLO'}:    & complete one-loop squared amplitude, \\
\texttt{'NLO-D4'}: & bare 4-dimensional one-loop squared amplitude, \\
\texttt{'NLO-CT'}: & counterterm contribution to the \\
                   & one-loop squared amplitude,\\
\texttt{'NLO-R2'}: & rational-term contribution to the\\
                   & one-loop squared amplitude.
\end{tabular}
\end{insertion}
\eei 
The argument \texttt{A2} of type {\tt real(dp)} delivers the
output consisting of the value of the squared amplitudes from the last
call of \texttt{compute\_process\_rcl} or
\texttt{rescale\_process\_rcl} for the process with process number
\npr.

\subsubsection{\tt{get\_polarized\_squared\_amplitude\_rcl (npr,pow,order,hel,A2h)}}

{\sloppy 
  This subroutine extracts the computed value for the contribution
  $\overline{{\cal A}_h^2}$ corresponding to the
  polarization $h$ = \hel\ to the squared amplitude for the process
  with identifier \npr\ (of type {\tt integer}), according to the
  values of the arguments \pow\ and \order.  The definition of the
  arguments \pow\ and \order\ is described in \refse{get squared},
  while the definition of \hel\ is described in \refse{get amplitude}.

The argument \texttt{A2h} of type {\tt real(dp)} delivers the
output consisting of the value of the polarized squared amplitude
from the last call of
\texttt{compute\_process\_rcl} or \texttt{rescale\_process\_rcl}
for the process with process number \npr.
}

\subsubsection{\tt{compute\_colour\_correlation\_rcl (npr,p,i1,i2,A2cc)}}

{\sloppypar
This subroutine computes the LO amplitudes 
${\cal A}_0^{(\vec{c},\vec{h})}$ and the specific 
colour-correlated squared amplitudes 
$\big(\,\overline{\!{\cal A}^2\!}\,\big)_{\!\mathrm{c}}(i_1,i_2)$ for 
the pair $(i_1,i_2)$ of external particles (see \refse{correlation} 
for details) for the process with identifier \npr\ 
(of type {\tt integer}). 
The {\tt integer} arguments \texttt{i1} and \texttt{i2}
denote the position identifiers of the respective external particles
(ordered as in the process definition). In the computation of the
squared amplitude a sum/average is performed over helicities and
colours (see \refse{squared amplitudes} for details on the definition
of squared amplitudes); for particles that have been defined with a
specific helicity, no helicity sum/average is performed.
}

The definition of the input argument \texttt{p(0:3,1:}$l$\texttt{)} of
type {\tt real(dp)} is described in \refse{compute-process}.

The results for the LO amplitudes and the colour-correlated squared
amplitudes are stored in internal variables (overwriting previously
computed results for LO amplitudes and colour-correlated squared
amplitudes for process \npr\ and particle pair \texttt{(i1,i2)}).  The
values of the colour-correlated squared amplitudes
$\big(\,\overline{\!{\cal A}^2\!}\,\big)_{\!\mathrm{c}}(i_1,i_2)$ can
be read out by the user with the subroutine
\texttt{get\_colour\_correlation\_rcl}.  The optional output argument
\texttt{A2cc} (of type {\tt real(dp)}) returns the value for the
colour-correlated squared amplitude, with the contributions from all
selected powers of $\alpha_s$ summed up.

\subsubsection{\tt{compute\_all\_colour\_correlations\_rcl (npr,p)}}

{\sloppy
This subroutine computes the LO amplitudes ${\cal A}_0^{(\vec{c},\vec{h})}$ 
and the colour-correlated squared amplitudes
$\big(\,\overline{\!{\cal A}^2\!}\,\big)_{\!\mathrm{c}}(i,j)$ for all
pairs $(i,j)$ of external coloured particles (see \refse{correlation} for
details) for the process with identifier \npr\ (of
type {\tt integer}).  In the computation of the squared amplitude a
sum/average is performed over helicities and colours (see
\refse{squared amplitudes} for details on the definition of squared
amplitudes); 
for particles that have been defined with a specific helicity, 
no helicity sum/average is performed.

The definition of input argument \texttt{p(0:3,1:}$l$\texttt{)} of
type {\tt real(dp)} is described in \refse{compute-process}.

The results for the LO amplitudes and the colour-correlated squared 
amplitudes are stored in internal variables (overwriting previously 
computed results for LO amplitudes and colour-correlated squared 
amplitudes for process \npr).
The values of the colour-correlated squared amplitudes 
$\big(\,\overline{\!{\cal A}^2\!}\,\big)_{\!\mathrm{c}}(i,j)$ can be 
read out by the user with the subroutine 
\texttt{get\_colour\_correlation\_rcl}.  
}
 
\subsubsection{\tt{rescale\_colour\_correlation\_rcl (npr,i1,i2,A2cc)}}

This subroutine can be employed to rescale the results calculated by
the subroutines \texttt{compute\_all\_colour\_correlations\_rcl} or
\texttt{compute\_colour\_correlation\_rcl} to a different value of
$\alpha_s$. To this end, the user first sets a new value for
$\alpha_s$ (by means of \texttt{set\_alphas\_rcl} or
\texttt{compute\_running\_alphas\_rcl}) and then calls
\texttt{rescale\_colour\_correlation\_rcl} with a process identifier
\npr\ and position identifiers \texttt{i1} and \texttt{i2} for the
respective external particles (\npr, \texttt{i1}, and \texttt{i2} are
of type {\tt integer}).  This leads to a rescaling of the stored
results for the LO amplitudes ${\cal A}_0^{(\vec{c},\vec{h})}$ and for
the colour-correlated squared amplitudes $\big(\,\overline{\!{\cal
A}^2\!}\,\big)_{\!\mathrm{c}}(i_1,i_2)$ for the process \npr\ and the
pair \texttt{(i1,i2)} of coloured external particles, overwriting
previous results.  The rescaled values of the colour-correlated
squared amplitudes $\big(\,\overline{\!{\cal
A}^2\!}\,\big)_{\!\mathrm{c}}(i_1,i_2)$ can be either obtained with
help of the subroutine \texttt{get\_colour\_correlation\_rcl} or via
the optional output argument \texttt{A2cc} (defined as for
\texttt{compute\_colour\_correlation\_rcl}).

\subsubsection{\tt{rescale\_all\_colour\_correlations\_rcl (npr)}}

This subroutine can be used to rescale the results calculated by the
subroutine \texttt{compute\_all\_colour\_correlations\_rcl} or
\texttt{compute\_colour\_correlation\_rcl} to a different value of
$\alpha_s$. To this end, the user first sets a new value for
$\alpha_s$ (by means of \texttt{set\_alphas\_rcl} or
\texttt{compute\_running\_alphas\_rcl}) and then calls
\texttt{rescale\_all\_colour\_correlations\_rcl} with a process
identifier \npr\ (of type {\tt integer}).
This leads to a rescaling of the stored results for the LO amplitudes
${\cal A}_0^{(\vec{c},\vec{h})}$ and for the colour-correlated squared
amplitudes $\big(\,\overline{\!{\cal
    A}^2\!}\,\big)_{\!\mathrm{c}}(i,j)$ for the process \npr\ and all
pairs $(i,j)$ of external coloured particles, overwriting previous
results.
The rescaled values of the colour-correlated squared amplitudes 
$\big(\,\overline{\!{\cal A}^2\!}\,\big)_{\!\mathrm{c}}(i,j)$ can be obtained
with help of the subroutine \texttt{get\_colour\_correlation\_rcl}.

\subsubsection{\tt{get\_colour\_correlation\_rcl (npr,pow,i1,i2,A2cc)}}

This subroutine extracts the computed value of the LO
colour-correlated squared amplitude $\big(\,\overline{\!{\cal
    A}^2\!}\,\big)_{\!\mathrm{c}}(i_1,i_2)$ for the pair $(i_1,i_2)$
of external particles (see \refse{correlation} for details) for the
process with identifier \npr\ (of type {\tt integer}).  The {\tt
  integer} arguments \texttt{i1} and \texttt{i2} denote the position
identifiers of the respective external particles (ordered as in the
process definition).  The {\tt integer} argument \pow\ specifies the
power of $\alpha_s$ of the contribution to be extracted.

{\sloppy
The argument \texttt{A2cc} of type {\tt real(dp)} delivers the output
of the subroutine consisting of the value of the LO colour-correlated
squared amplitudes for particle pair \texttt{(i1,i2)}
from the last call of \texttt{compute\_colour\_correlation\_rcl} or
\texttt{rescale\_colour\_correlation\_rcl}
for the process with process number \npr\
and external coloured particles \texttt{(i1,i2)},
or from the last call of \texttt{compute\_all\_colour\_correlations\_rcl} or
\texttt{rescale\_all\_colour\_correlations\_rcl}
for the process with process number \npr.}

\subsubsection{\tt{compute\_spin\_correlation\_rcl 
                       (npr,p,j,v,A2sc)}}
\label{compute spin}

This subroutine computes the LO amplitudes ${\cal
A}_0^{(\vec{c},\vec{h})}$ and the spin-correlated squared amplitude
$\big(\,\overline{\!{\cal A}^2\!}\,\big)_{\!\mathrm{s}}(j)$ for the
photon $j$ in the process with identifier \npr\ (of type {\tt
integer}), using as polarization vector for the photon $j$ the
four-vector \texttt{v} provided by the user (see \refse{correlation}
for details).  The {\tt integer} argument \texttt{j} denotes the
position identifier of the respective photon (following the order of
external particles in the process definition).  The vector
\texttt{v(0:3)} (of type {\tt complex(dp)}) denotes the user-defined
four-vector substituting the polarization vector of photon $j$.  In
the computation of the squared amplitude a sum/average is performed
over helicities and colours (see \refse{squared amplitudes} for
details on the definition of squared amplitudes); for particles that
have been defined with a specific helicity as well as for the photon
$j$, no helicity sum/average is performed.

The results for the LO amplitudes and the spin-correlated squared
amplitudes are stored in internal variables (overwriting previously
computed results for LO amplitudes and spin-correlated squared
amplitudes for process \npr).  The values of the spin-correlated
squared amplitudes $\big(\,\overline{\!{\cal
    A}^2\!}\,\big)_{\!\mathrm{s}}(j)$ can be read out by the user with
the subroutine \texttt{get\_spin\_correlation\_rcl}.  The optional
output argument \texttt{A2sc} (of type {\tt real(dp)}) returns the
value for the spin-correlated squared amplitude with the contributions
from all selected powers of $\alpha_s$ summed up.

\subsubsection{\tt{rescale\_spin\_correlation\_rcl (npr,j,v,A2sc)}}

This subroutine allows the user to rescale the results calculated by
\texttt{compute\_spin\_correlation\_rcl} to a different value of
$\alpha_s$. To this end, the user first sets a new value for
$\alpha_s$ (by means of \texttt{set\_alphas\_rcl} or
\texttt{compute\_running\_alphas\_rcl}) and then calls
\texttt{rescale\_spin\_correlation\_rcl} with a process identifier
\npr\ and an identifier \texttt{j} for the position of the external
photon (\npr\ and \texttt{j} are of type {\tt integer}).  This leads
to a rescaling of the stored results for the LO amplitudes ${\cal
A}_0^{(\vec{c},\vec{h})}$ and for the spin-correlated squared
amplitudes $\big(\,\overline{\!{\cal A}^2\!}\,\big)_{\!\mathrm{s}}(j)$
for the process \npr\ and the photon \texttt{j}, overwriting previous
results.  The vector \texttt{v(0:3)} (of type {\tt complex(dp)}) plays
the same role as in \texttt{compute\_spin\_correlation\_rcl}.  While
the rescaling of the LO amplitudes ${\cal A}_0^{(\vec{c},\vec{h})}$ is
not affected by the value of \texttt{v}, the spin-correlated squared
amplitudes $\big(\,\overline{\!{\cal A}^2\!}\,\big)_{\!\mathrm{s}}(j)$
are computed by \texttt{rescale\_spin\_correlation\_rcl} for the
current value of \texttt{v}, independently of previous computations of
spin correlations.  The rescaled result can be either obtained with
help of the subroutine \texttt{get\_spin\_correlation\_rcl} or via the
optional output argument \texttt{A2sc} (defined as for
\texttt{compute\_spin\_correlation\_rcl}).

\subsubsection{\tt{get\_spin\_correlation\_rcl (npr,pow,A2sc)}}

This subroutine extracts the computed value of the LO spin-correlated
squared amplitude $\big(\,\overline{\!{\cal
    A}^2\!}\,\big)_{\!\mathrm{s}}$ (see \refse{correlation} for
details) for the process with identifier \npr\ (of type {\tt
  integer}).  The {\tt integer} argument \pow\ specifies the power of
$\alpha_s$ of the contribution to be extracted.

The argument \texttt{A2sc} of type {\tt real(dp)} delivers the output
of the subroutine consisting of the value of the LO spin-correlated
squared amplitudes from the last call of
\texttt{compute\_spin\_correlation\_rcl} or
\texttt{rescale\_spin\_correlation\_rcl} for the process with process
number \npr.

\subsubsection{\tt{compute\_spin\_colour\_correlation\_rcl 
                       (npr,p,i1,i2,v,A2scc)}}

This subroutine computes the LO amplitudes ${\cal
A}_0^{(\vec{c},\vec{h})}$ and the specific colour-correlated squared
amplitude $\big(\,\overline{\!{\cal
A}^2\!}\,\big)_{\!\mathrm{sc}}(i_1,i_2)$ for the pair $(i_1,i_2)$ of
an external gluon $i_1$ and a spectator $i_2$ (see \refse{correlation}
for details) for the process with identifier \npr\ (of type {\tt
integer}).  The {\tt integer} arguments \texttt{i1} and \texttt{i2}
denote the position identifiers of the respective external particles
(ordered as in the process definition).  The polarization vector of
the gluon $i_1$ is substituted by the four-vector \texttt{v(0:3)} (of
type {\tt complex(dp)}) provided by the user (see \refse{correlation}
for details).  In the computation of the squared amplitude a
sum/average is performed over helicities and colours (see
\refse{squared amplitudes} for details on the definition of squared
amplitudes); for particles that have been defined with a specific
helicity and for the gluon $i_1$, no helicity sum/average is
performed.

The definition of input argument \texttt{p(0:3,1:}$l$\texttt{)} of
type {\tt real(dp)} is described in \refse{compute-process}.

The results for the LO amplitudes and the spin--colour-correlated squared 
amplitudes are stored in internal variables (overwriting previously
computed results for LO amplitudes and spin--colour-correlated squared
amplitudes for process \npr\
and particle pair \texttt{(i1,i2)}).  The
values of the spin--colour-correlated squared amplitudes
$\big(\,\overline{\!{\cal A}^2\!}\,\big)_{\!\mathrm{sc}}(i_1,i_2)$ can
be read out by the user with the subroutine
\texttt{get\_spin\_colour\_correlation\_rcl}.  The optional output
argument \texttt{A2scc} (of type {\tt real(dp)}) returns the value for
the spin--colour-correlated squared amplitude for particle pair
\texttt{(i1,i2)} with the contributions from all selected powers of
$\alpha_s$ summed up.

\subsubsection{\tt{rescale\_spin\_colour\_correlation\_rcl (npr,i1,i2,v,A2scc)}}

This subroutine can be used to rescale the results calculated by
\texttt{compute\_spin\_colour\_correlation\_rcl} to a different value
of $\alpha_s$. To this end, the user first sets a new value for
$\alpha_s$ (by means of \texttt{set\_alphas\_rcl} or
\texttt{compute\_running\_alphas\_rcl}) and then calls
\texttt{rescale\_spin\_colour\_correlation\_rcl} with a process
identifier \npr\ and position identifiers \texttt{i1} for the gluon
and \texttt{i2} for the spectator (\npr, \texttt{i1}, and \texttt{i2}
are of type {\tt integer}).
This leads to a rescaling of the stored results for the LO amplitudes
${\cal A}_0^{(\vec{c},\vec{h})}$ and for the spin--colour-correlated
squared amplitudes $\big(\,\overline{\!{\cal
A}^2\!}\,\big)_{\!\mathrm{sc}}(i_1,i_2)$ for the process \npr\ and the
particle pair \texttt{(i1,i2)}, overwriting previous results.  The
vector \texttt{v(0:3)} (of type {\tt complex(dp)}) plays the same role
as in \texttt{compute\_spin\_colour\_correlation\_rcl}.  While the
rescaling of the LO amplitudes ${\cal A}_0^{(\vec{c},\vec{h})}$ is not
affected by the value of \texttt{v}, the spin--colour-correlated
squared amplitudes $\big(\,\overline{\!{\cal
A}^2\!}\,\big)_{\!\mathrm{sc}}(i_1,i_2)$ are computed by
\texttt{rescale\_spin\_correlation\_rcl} for the current value of
\texttt{v}, independently of previous computations of spin--colour
correlations.
The rescaled result can be either obtained
with help of the subroutine \texttt{get\_spin\_colour\_correlation\_rcl}
or via the optional output 
argument \texttt{A2scc} (defined as for \texttt{compute\_spin\_colour\_correlation\_rcl}).

\subsubsection{\tt{get\_spin\_colour\_correlation\_rcl 
                       (npr,pow,i1,i2,A2scc)}}

This subroutine extracts the computed value of the LO 
spin--colour-correlated squared amplitude 
$\big(\,\overline{\!{\cal A}^2\!}\,\big)_{\!\mathrm{sc}}(i_1,i_2)$ for the 
pair $(i_1,i_2)$ of gluon $i_1$ 
and spectator $i_2$ (see \refse{correlation} for
details) for the process with identifier \npr\ (of type {\tt integer}).
The {\tt integer} arguments \texttt{i1} and \texttt{i2}
denote the position 
identifiers of the respective external particles
(ordered as in the process definition).
The {\tt integer} argument \pow\ specifies the power of $\alpha_s$ of the 
contribution to be extracted.
The argument \texttt{A2scc} of type {\tt real(dp)} delivers the output
of the subroutine consisting of the value of the LO
spin--colour-correlated squared amplitudes as stored in the internal
variables by the last call of
\texttt{compute\_spin\_colour\_correlation\_rcl}, or
\texttt{rescale\_spin\_colour\_correlation\_rcl} for the process with
process number \npr\ and particle pair \texttt{(i1,i2)}.

\subsubsection{\tt{get\_momenta\_rcl (npr,p)}}

This subroutine extracts the momenta of the process with 
identifier \npr\ (of type {\tt integer}), stored
from the last call of a subroutine of type \texttt{compute\_...\_rcl}
for process \npr. The output array \texttt{p} is of type {\tt real(dp)}
and has the format \texttt{p(0:3,1:$l$)},
where $l$ is the number of external particles of
process \npr.

Note that \recola\ adjusts the momenta provided by the user if it
detects violation of momentum conservation or mass-shell conditions.

\subsubsection{\tt{set\_TIs\_required\_accuracy\_rcl (acc)}}

This subroutine sets the required accuracy for TIs to the value
\texttt{acc} (of type {\tt real(dp)}).  This parameter is passed to
\collier\ as target accuracy $\eta_{\text req}$ (see
\citere{Denner:2016kdg} for details).

\subsubsection{\tt{get\_TIs\_required\_accuracy\_rcl (acc)}}

This subroutine extracts the value of the required accuracy
$\eta_{\text req}$ for TIs from \collier\ (see \citere{Denner:2016kdg}
for details) and returns it as value of the output variable
\texttt{acc} (of type {\tt real(dp)}).

\subsubsection{\tt{set\_TIs\_critical\_accuracy\_rcl (acc)}}

This subroutine sets the critical accuracy for TIs to the value
\texttt{acc} (of type {\tt real(dp)}).  This parameter is passed to
\collier\ as critical accuracy $\eta_{\text crit}$ (see
\citere{Denner:2016kdg} for details).

\subsubsection{\tt{get\_TIs\_critical\_accuracy\_rcl (acc)}}

This subroutine extracts the value of the critical accuracy
$\eta_{\text crit}$ for TIs from \collier\ (see
\citere{Denner:2016kdg} for details) and returns it as value of the
output variable \texttt{acc} (of type {\tt real(dp)}).

\subsubsection{\tt{get\_TIs\_accuracy\_flag\_rcl (flag)}}

This subroutine extracts the value of the accuracy flag $\sigma_{\text
acc}$ for TIs from \collier\ (see \citere{Denner:2016kdg} for
details).  The output variable \texttt{flag} (of type {\tt integer})
returns global information on the accuracy of the TIs evaluated in the
last call of \texttt{compute\_process\_rcl}:
\bei
\item
\texttt{flag = 0}: For all TIs, the accuracy is estimated to be better 
than the required value.
\item
\texttt{flag = -1}: For at least one TI, the accuracy is estimated to be
worse than the required value, but for all TIs, 
the accuracy is estimated to be 
better than the critical value.
\item
\texttt{flag = -2}: For at least one TI, the accuracy is estimated to be
worse than the critical values.
\eei
The value of variable \texttt{flag} is determined based on internal
uncertainty estimations performed by \collier.


\subsection{Reset: \tt{reset\_recola\_rcl}}
\label{reset}

The file \texttt{reset.f90} only contains the single subroutine
\texttt{reset\_recola\_rcl} which can be called to free memory and to
allow for the definition of a new set of processes in the same run of
the program.  A call of this subroutine deallocates all global
allocatable arrays internally generated by \recola\ and restores the
initialization values for a bunch of internal variables, allowing to
restart the application of \recola\ with step 1 or 2 of the sequence
defined at the beginning of \refse{calling recola}.  The input
variables in \texttt{input.f90} keep their actual values.  The call of
\texttt{reset\_recola\_rcl} resets the name of the output file to the
default value \texttt{output.rcl}.

\section{Conclusions}
\label{conclusions}

The {\sc Fortran}-based library
\recola\ calculates amplitudes and squared
amplitudes in the Standard Model of particle physics including QCD and
the electroweak interaction at the tree and one-loop level with no
a-priori restriction on the particle multiplicities. Amplitudes can be
obtained for specific colour structures and helicities and squared
amplitudes with or without summation/average over helicities.
Moreover, colour- and spin-correlated leading-order squared amplitudes
for dipole subtraction are provided.

Renormalization is performed in the complex-mass scheme, or
alternatively in the on-shell scheme, and various renormalization
schemes are supported for the electromagnetic coupling.  For the
strong coupling, fixed or dynamical $N_f$-flavour schemes are
available.  Infrared singularites can be regularized dimensionally or
with infinitesimal fermion and photon/gluon masses. The code allows to
select contributions involving specific resonances.

The present version of the code is restricted to the Standard Model in
the 't Hooft--Feynman gauge. A version for more general theories is in
preparation.

\section{Acknowledgements}
We thank B.~Biedermann, R.~Feger, and M.~Pellen for performing various
checks of the code.  This work was supported by the Deutsche
Forschungsgemeinschaft (DFG) under reference number DE~623/2-1.  The
work of L.H.\ was supported by the grants FPA2013-46570-C2-1-P and
2014-SGR-104, and partially by the Spanish MINECO under the project
MDM-2014-0369 of ICCUB (Unidad de Excelencia ``Mar\'ia de Maeztu'').
The work of S.U.\ was supported in part by the European Commission through
the ‘HiggsTools’ Initial Training Network PITN-GA-2012-316704.

\appendix

\section{Explicit representations for spinors and polarization vectors}
\label{appendix_spinors}

Here we list 
the explicit expressions of the spinors and polarization 
vectors used in \recola, which are 
in the chiral representation for the Dirac matrices:
\begin{itemize}
\item Spinors for massive fermions:
$$
u_+(p) = 
\frac{1}{r}
\left(
\begin{array}{c}
a_+b_+\\
s a_+b_-\hat{p}_+\\
-s a_-b_+\\
-a_-b_-\,\hat{p}_+
\end{array} 
\right),
\quad
u_-(p) =
\frac{1}{r}
\left(
\begin{array}{c}
a_-b_-\hat{p}_-\\
-s a_-b_+\\
-s a_+b_-\hat{p}_-\\
a_+b_+
\end{array} 
\right),
$$
$$
v_+(p) =
\frac{1}{r}
\left(
\begin{array}{c}
-a_-b_-\hat{p}_-\\
s a_-b_+\\
-s a_+b_-\hat{p}_-\\
a_+b_+
\end{array} 
\right),
\quad
v_-(p) = 
\frac{1}{r}
\left(
\begin{array}{c}
a_+b_+\\
s a_+b_-\hat{p}_+\\
s a_-b_+\\
a_-b_-\hat{p}_+
\end{array} 
\right),
$$
\bqa
\bar{u}_+(p)
&=&
\frac{1}{r}\,
\Big(
s a_-b_+,\;
a_-b_-\hat{p}_-,\;
-a_+b_+,\;
-s a_+b_-\hat{p}_-
\Big),
\nonumber\\
\bar{u}_-(p)
&=& 
\frac{1}{r}\,
\Big(
s a_+b_-\hat{p}_+,\;
-a_+b_+,\;
-a_-b_-\hat{p}_+,\;
s a_-b_+
\Big),
\nonumber\\
\bar{v}_+(p) 
&=& 
\frac{1}{r}\,
\Big(
s a_+b_-\hat{p}_+,\;
-a_+b_+,\;
a_-b_-\hat{p}_+,\;
-s a_-b_+
\Big),
\nonumber\\
\bar{v}_-(p) 
&=&
\frac{1}{r}\,
\Big(
-s a_-b_+,\;
-a_-b_-\hat{p}_-,\;
-a_+b_+,\;
-s a_+b_-\hat{p}_-
\Big),
\eqa
with
$$
r = \sqrt{2|\vec{p}|},
\qquad\!\!\!
a_\pm = \sqrt{|p_0| \pm |\vec{p}|},
\qquad\!\!\!
b_\pm = \sqrt{|\vec{p}| \pm s\,p_z},
\qquad\!\!\!
s = {\rm sign}(p_0),
$$
\bq
\hat{p}_\pm = \frac{p_\pm}{p_{\!_T}},
\qquad
p_\pm = p_x \pm {\rm i}\,p_y,
\qquad
p_{\!_T} = \sqrt{p_x^2 + p_y^2},
\qquad
|\vec{p}| = \sqrt{p_{\!_T}^2 + p_z^2}.
\eq
\item Spinors for massless fermions:
$$
u_+(p) = v_-(p) = 
\left(
\begin{array}{c}
b_+\\
s b_-\hat{p}_+\\
0 \\
0
\end{array} \right),
\qquad
u_-(p) = v_+(p) =
\left(
\begin{array}{c}
0 \\
0 \\
-s b_-\hat{p}_-\\
b_+
\end{array} 
\right),
$$
$$
\bar{u}_+(p) 
= 
\bar{v}_-(p) 
=
\Big(
0,\;
0,\;
-b_+,\;
-s b_-\hat{p}_-
\Big),
$$
\bq
\bar{u}_-(p) 
=
\bar{v}_+(p) 
=
\Big(
s b_-\hat{p}_+,\;
-b_+,\;
0,\;
0
\Big).
\eq
\item Transverse polarization vectors (for massless and massive vector bosons):
\bq
\epsilon_\pm = 
\frac{1}{\sqrt{2}\,|\vec{p}|\,p_T}\,
\Big(\,
0, \;
\mp s\,p_x\,p_z + {\rm i}\,p_y\,|\vec{p}|,\;
\mp s\,p_y\,p_z - {\rm i}\,p_x\,|\vec{p}|,\;
\pm s\,p_T^2
\,\Big).
\eq

\item Longitudinal polarization vectors (for massive vector bosons with mass $M$):
\bq
\epsilon_0 = 
\frac{1}{M\,|\vec{p}|}
\Big(\,
|\vec{p}|^2,\;
p_x\,p_0,\;
p_y\,p_0,\;
p_z\,p_0
\,\Big).
\eq
\end{itemize}

\section{Checks}
\label{checks}

A variety of processes has been checked at LO and NLO with
the in-house code \pole~\cite{Accomando:2005ra} and the code 
\openloops~\cite{Cascioli:2011va}. 
The typical agreement for NLO matrix elements ranges 
from $10^{-14}$ for processes with 4 external legs to $10^{-10}$ for processes
with 6 external legs in the comparison with \pole, and from $10^{-12}$
to $10^{-8}$ in the comparison with \openloops.  Note that the
precision of the NLO matrix element strongly depends on the
phase-space point for which it is evaluated, and is limited by the
accuracy of the tensor intagrals, compared to which numerical
uncertainties from the tensor coefficients calculated by \recola\ are
completely negligible.

The following processes have been checked with the code \pole\ 
in a Monte Carlo integration for the calculations in \citeres{Actis:2012qn,Denner:2014ina,Biedermann:2016yvs}:
\bei
\item
5-leg processes (all powers of $\alpha_s$ at LO, ${\cal O}(\alpha_s^2\alpha)$ and 
${\cal O}(\alpha_s\alpha^2)$ at NLO):\\
${\rm u} \, {\rm g}     \to {\rm u}   \, {\rm g} \, {\rm Z}$ \\
${\rm d} \, {\rm g}     \to {\rm d}   \, {\rm g} \, {\rm Z}$ \;,

\item 6-leg process (all powers of $\alpha_s$ at LO, ${\cal
    O}(\alpha_s^2\alpha)$ and
  ${\cal O}(\alpha_s\alpha^2)$ at NLO):\\
${\rm u} \, {\rm d} \to {\rm u} \, {\rm d} \, {\rm e}^+ \, {\rm e}^-$\;, 

\item
6-leg process (all powers of $\alpha_s$ at LO and NLO):\\
${\rm u} \, \bar{\rm u} \to \mu^+ \, \mu^-\,  {\rm e}^+ \, {\rm e}^-$\;.
\eei
The following processes have been checked for single phase-space points: 
\bei
\item
4-leg processes (EW) checked with the code \pole:\\
$\bar{\rm u} \, {\rm u}     \to \bar{\nu}_e \, \nu_e$ \\
${\rm u}     \, \bar{\rm d} \to \nu_e       \, {\rm e}^+$ \\
${\rm e}^+   \, {\rm e}^-   \to \bar{\nu}_e \, \nu_e$ \\
${\rm e}^+   \, {\rm e}^-   \to {\rm W}^+   \, {\rm W}^-$ \\
${\rm e}^+   \, {\rm e}^-   \to {\rm Z}     \, {\rm H}$ \;,

\item
4-leg processes (QCD) checked with the code \openloops:\\
${\rm u} \, \bar{\rm d} \to {\rm W}^+ \, {\rm g}$ \\
${\rm g} \, {\rm g}     \to {\rm g}   \, {\rm g}$ \\
${\rm b} \, \bar{\rm b} \to {\rm t}   \, \bar{\rm t}$ \\
${\rm g} \, {\rm g}     \to {\rm b}   \, \bar{\rm b}$ \;,
\item
4-leg processes (EW+QCD) checked with the code \pole:\\
$\bar{\rm d} \, {\rm d}     \to \bar{\rm u} \, {\rm u}$ \\
${\rm u}     \, \bar{\rm d} \to {\rm W}^+   \,  {\rm H}$ \;,
\item
5-leg processes (EW) checked with the code \pole:\\
${\rm u} \, \bar{\rm d} \to {\rm e}^+ \, \nu_e   \, \gamma$ \;,
\item
5-leg processes (QCD) checked with the code \openloops:\\
${\rm u} \, \bar{\rm u} \to {\rm W}^+ \, {\rm W}^-    \, {\rm g}$ \\
${\rm u} \, \bar{\rm u} \to {\rm Z}   \, {\rm Z}     \, {\rm g}$ \\
${\rm u} \, \bar{\rm u} \to {\rm Z}   \, \gamma      \, {\rm g}$ \\
${\rm u} \, \bar{\rm u} \to \gamma    \, \gamma      \, {\rm g}$ \\
${\rm u} \, \bar{\rm d} \to {\rm W}^+ \, {\rm g}     \, {\rm g}$ \\
${\rm u} \, \bar{\rm d} \to {\rm W}^+ \, {\rm t}     \, \bar{\rm t}$ \\
${\rm u} \, \bar{\rm u} \to {\rm Z}   \, {\rm t}     \, \bar{\rm t}$ \\
${\rm d} \, \bar{\rm d} \to {\rm Z}   \, {\rm t}     \, \bar{\rm t}$ \\
${\rm g} \, {\rm g}     \to {\rm W}^+ \, {\rm b}     \, \bar{\rm t}$ \\
${\rm g} \, {\rm g}     \to {\rm Z}   \, {\rm t}     \, \bar{\rm t}$ \\
${\rm u} \, \bar{\rm u} \to {\rm Z}   \, {\rm g}     \, {\rm g}$ \\
${\rm g} \, {\rm g}     \to {\rm g}   \, {\rm t}     \, \bar{\rm t}$ \\
${\rm g} \, {\rm g}     \to {\rm g}   \, {\rm g}     \, {\rm g}$ \\
${\rm d} \, \bar{\rm d} \to {\rm d}   \, \bar{\rm d} \, {\rm g}$ \\
${\rm d} \, \bar{\rm d} \to {\rm t}   \, \bar{\rm t} \, {\rm g}$ \\
${\rm b} \, \bar{\rm b} \to {\rm t}   \, \bar{\rm t} \, {\rm g}$ \;,
\item
6-leg processes (QCD) checked with the code \openloops:\\
${\rm u} \, \bar{\rm d} \to {\rm W}^+ \, {\rm g}     \, {\rm g} \, {\rm g}$ \\
${\rm u} \, \bar{\rm u} \to {\rm Z}   \, {\rm g}     \, {\rm g} \, {\rm g}$ \\
${\rm u} \, \bar{\rm u} \to {\rm W}^+ \, {\rm W}^-    \, {\rm g} \, {\rm g}$ \\
${\rm u} \, \bar{\rm u} \to {\rm Z}   \, {\rm Z}     \, {\rm g} \, {\rm g}$ \\
${\rm d} \, \bar{\rm d} \to {\rm t}   \, \bar{\rm t} \, {\rm b} \, \bar{\rm b}$ \\
${\rm g} \, {\rm g}     \to {\rm t}   \, \bar{\rm t} \, {\rm b} \, \bar{\rm b}$ \\
${\rm u} \, \bar{\rm u} \to {\rm u}   \, \bar{\rm u} \, {\rm u} \, \bar{\rm u}$ \\
${\rm g} \, {\rm g}     \to {\rm u}   \, \bar{\rm u} \, {\rm u} \, \bar{\rm u}$ \\
${\rm g} \, {\rm g}     \to {\rm u}   \, \bar{\rm u} \, {\rm d} \, \bar{\rm d}$ \\
${\rm g} \, {\rm g}     \to {\rm u}   \, \bar{\rm u} \, {\rm g} \, {\rm g}$ \\
${\rm g} \, {\rm g}     \to {\rm t}   \, \bar{\rm t} \, {\rm g} \, {\rm g}$ \;.
\eei
Thereby (EW) refers to contributions to the squared amplitudes with
minimal power of $\alpha_\mathrm{s}$ at LO and at NLO, 
(QCD) refers to contributions to the squared amplitudes with 
maximal power of $\alpha_s$ at LO and at NLO and 
(EW+QCD) refers to the sum of all contributions to the squared 
amplitudes at LO and at NLO.



\bibliographystyle{elsarticle-num}
\bibliography{recola}


\end{document}